\documentclass[aps,prb,floatfix,onecolumn,superscriptaddress,nofootinbib]{revtex4}

\usepackage{cmbright}
\DeclareFontShape{OT1}{cmss}{m}{it}{<->ssub*cmss/m/sl}{}

\usepackage{amsmath,color,graphicx}
\usepackage[utf8]{inputenc}
\usepackage[T1]{fontenc}
\graphicspath{{../figures/}}

\begin{document}

\title{Field-induced vortex-like textures as a probe of the critical line in reentrant spin glasses}
\author{N. Martin}\email[]{nicolas.martin@cea.fr}\affiliation{Universit\'e Paris-Saclay, CEA, CNRS, Laboratoire L\'eon Brillouin,  CEA Saclay 91191 Gif-sur-Yvette, France}
\author{L.J. Bannenberg}\affiliation{Faculty of Applied Science, Delft University of Technology, 2629 JB Delft, the Netherlands}
\author{M. Deutsch}\affiliation{Universit\'e de Lorraine, CNRS, CRM2, Nancy, France}
\author{C. Pappas}\affiliation{Faculty of Applied Science, Delft University of Technology, 2629 JB Delft, the Netherlands}
\author{G. Chaboussant}\affiliation{Universit\'e Paris-Saclay, CEA, CNRS, Laboratoire L\'eon Brillouin,  CEA Saclay 91191 Gif-sur-Yvette, France}
\author{R. Cubitt}\affiliation{Institut Laue Langevin, BP156, F-38042 Grenoble France}
\author{I. Mirebeau}\affiliation{Universit\'e Paris-Saclay, CEA, CNRS, Laboratoire L\'eon Brillouin,  CEA Saclay 91191 Gif-sur-Yvette, France}

\date{\today}

\begin{abstract}
We study the evolution of the low-temperature field-induced magnetic defects observed under an applied magnetic field in a series of frustrated amorphous ferromagnets (Fe$_{1-x}$Mn$_{x}$)$_{75}$P$_{16}$B$_{3}$Al$_{3}$ (a-FeMn). Combining small-angle neutron scattering and Monte Carlo simulations, we show that the morphology of these defects resemble that of quasi-bidimensional spin vortices. They are observed in the reentrant spin-glass (RSG) phase, up to the critical concentration $x_{\rm C} \approx 0.36$ which separates the RSG and "true" spin glass (SG) within the low temperature part of the magnetic phase diagram of a-FeMn. These vortices systematically decrease in size with increasing magnetic field or decreasing the average exchange interaction, and they finally disappear in the SG sample ($x = 0.41$), being replaced by field-induced correlations over finite length scales. We argue that the study of these nanoscopic defects could be used to probe the nature of the critical line between the RSG and SG phases. 
\end{abstract}

\maketitle

\section*{Introduction}
The role of disorder is central in condensed matter physics, as it favors the nucleation of defects which play a crucial role in the evolution and functionalities of a large variety of systems. Examples are magnetic vortices in superconductors, skyrmions in helical magnets, Taylor cells in liquid flows, or twist grain boundary phases in cholesteric liquid crystals. Quite generally, defects allow new properties to penetrate in the system by forming intermediate states of matter, precursors of a phase transitions. In this context, we study here the influence of nm-size magnetic defects on the evolution from ferromagnetic (FM) and spin glass (SG) ground states. SG are archetypal disordered magnetic systems that have mobilized a large and continuous attention for decades. Their physics is mainly driven by atomic disorder {\it and} random sign interactions, i.e. a mixture of FM and antiferromagnetic (AFM) couplings usually tuned by the concentration of magnetic ions. In the so-called reentrant spin glasses (RSG), the nature of the coexistence between SG behavior and ferromagnetism has been much debated. In this work, we show that the observation of magnetic field-induced vortices, although not predicted by current theories, could be a key point to distinguish between SG and RSG ground states at a microscopic level. 

Historically, the RSG and SG ground states have been described by two concurrent theoretical approaches. The infinite range mean field (MF) picture yields a phase diagram with a tricritical point and a vertical line between RSG and SG phases\cite{Gabay1981}. Below this line, {\it i.e.} in the weakly frustrated case, lowering temperature leads to the occurrence of mixed RSG phases, where SG and FM order parameters coexist microscopically. At each magnetic site, "longitudinal" spin component, forming a long-range magnetic order (LRMO), coexist with a "transverse" one, randomly oriented in the perpendicular plane. Alternatively, random field (RF) arguments predict the breakdown of LRMO for an arbitrarily small amount of disorder in dimensions $d \leq 4$, as formalized by the Imry-Ma (IM) argument \cite{Imry1975}. This argument was used together with percolation approaches to describe the RSG phase as randomly oriented clusters spatially separated from the "infinite" FM one. The latter would break due to RFs at the RSG-SG threshold, in a cross-over transition \cite{Aeppli1983,Niidera2007}. The IM argument was recently complemented by a series of Monte-Carlo (MC) simulations suggesting that, in the case of ferromagnets subject to RF, the IM domains are protected against a full collapse of the magnetization by the nucleation of topological defects, such as pinned hedgedhogs in 3 dimensions\cite{Proctor2014} or a "skyrmion-antiskyrmion glass" in 2 dimensions\cite{Chudnovsky2018}.

 On the experimental side, magnetic defects -akin to nm-size vortices- have been observed by small-angle neutron scattering (SANS) in weakly frustrated RSG under an applied magnetic field. The family of studied compounds (Ni$_{1-x}$Mn$_{x}$, Au$_{x}$Fe$_{1-x}$, Fe$_{1-x}$Al$_{x}$, or a-Fe$_{1-x}$Mn$_{x}$) includes different types of disorder, magnetic interactions and sample form  (single crystal, polycrystal or amorphous samples)\cite{Hennion1986,Boeni1986,Lequien1987,Hennion1988}. In all cases, SANS experiments show that the transverse spin components rotate over a finite length scale which defines the average vortex size. These data, supported by MC simulations\cite{Kawamura1991}, also indicate that the vortices shrink with increasing the applied field, but their behaviour at strong frustration and across the RSG-SG threshold has not been studied so far.

In order to address this point, we focus here on the series of frustrated amorphous ferromagnets (Fe$_{1-x}$Mn$_{x}$)$_{75}$P$_{16}$B$_6$Al$_3$ ("a-FeMn"). a-FeMn maps a case of 3d disordered Heisenberg spins, where frustration can be chemically tuned through the competition of FM and AFM interactions. Using SANS, we follow the evolution of the field-induced vortices with increasing frustration, as the magnetic ground state evolves from RSG to SG. We show that non-singular vortices are characteristic of the RSG ground state. Their average size obeys scaling laws up to the critical concentration and maximum applied field. They eventually disappear above the RSG-SG threshold, showing that a non-zero average exchange is needed for their stabilization. Our results may open a route to reconcile the MF and RF pictures of the RSG state, opposed for decades. Indeed, vortices can probe the nature of the frustrated medium and be the fingerprints of a quantum phase transition at a microscopic level. Our experimental study is also supported by Monte Carlo simulations, which show that the occurrence of these defects, as well as their evolution as a function of magnetic field, can be globally reproduced using a very limited amount of ingredients. These results suggest that frustration and disorder can be used to engineer the average size of individual defects in a controlled and reproducible way in disordered frustrated ferromagnets.

\section*{Samples and their macroscopic magnetic properties}

The a-FeMn system is perfectly suitable for our study. Its macroscopic properties and transition temperatures are well known and almost independent of the sample synthesis, while the amorphous character guarantees the absence of structural defects which could otherwise act as pinning centers\cite{Yeshurun1981}. Frustration is monitored by the Mn concentration $x$ which controls the relative amounts of AFM Mn-Mn nearest neighbor (NN) bonds with respect to the FM Mn-Fe and Fe-Fe ones. Here, we study seven RSG samples of concentrations ranging from $x = 0.22$ to $x = 0.35$ and a SG sample with $x = 0.41$, previously studied by neutron depolarization and muon spin rotation\cite{Mirebeau1997}. Samples were prepared using the "wheelbarrow" technique and their amorphous nature was checked using neutron diffraction (see Supplementary Information with Supplementary Figs. 1 and 2). The resulting ribbons, of typical thickness $\approx$ 30-70 $\mu$m and $\approx$ 8-10 mm width, can be easily cut or piled-up to perform magnetic and neutron scattering experiments. 

\begin{figure}[!ht]
	\includegraphics[width=\textwidth]{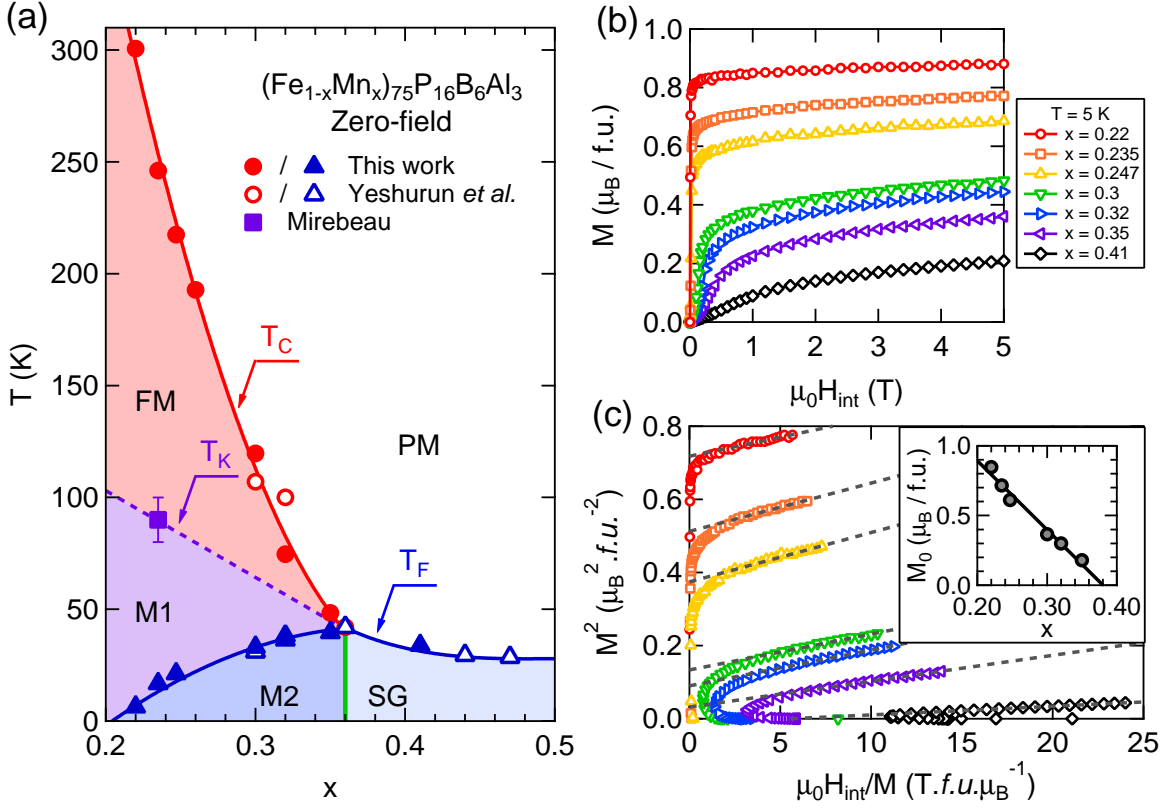}
	\caption{{\bf Phase diagram and magnetization of a-FeMn --} \textbf{(a)} Magnetic phase diagram of a-Fe$_{1-x}$Mn$_{x}$ inferred from magnetic ac-susceptibility. For $0.2 \lesssim x \lesssim x_{\rm C}$, cooling from the high-temperature FM state leads to a sequence of two mixed states: M1, involving the freezing of transverse spin components \cite{Mirebeau1990} and M2, where replica symmetry is spontaneously broken (i.e., an analog of the SG state). Above $x = x_{\rm C} \approx 0.36$, the ferromagnetic (FM) phase is suppressed and replaced with a "canonical" spin-glass (SG) state at low temperature. Data from Mirebeau\cite{Mirebeau1987} and Yeshurun {\it et al.}\cite{Yeshurun1981} is added for the sake of completeness. The green vertical line indicates a putative critical line between the RSG (M2 phase) and SG regimes\cite{Gabay1981}. \textbf{(b)} Low-temperature magnetization curves of a-FeMn. \textbf{(c)} Arrot plots computed from the data of panel \textbf{(b)}. Dashed lines are linear fits to the high-field data. Inset shows the $x$-dependence of the spontaneous magnetization $M_{\rm 0} = M(\mu_{0}H_{\rm int} \rightarrow 0)$.}
	\label{fig:diagram-mag}
\end{figure}

The magnetic phase diagram of a-FeMn, inferred from ac-susceptibility (see Supplementary Information with Supplementary Figs. 4, 5 and 6), is shown in Fig. \ref{fig:diagram-mag}a. As already found by Yeshurun {\it et al.}\cite{Yeshurun1981}, transition lines separating the paramagnetic-FM states ($T_{\rm C}$) and FM-RSG or paramagnetic-SG states ($T_{\rm F}$) merge at a tricritical point located at $x_{\rm C} \approx 0.36$ (as also confirmed by magnetic susceptibility\cite{Salamon1980}, as well as neutron depolarization\cite{Mirebeau1997} measurements). The Curie ($T_{\rm C}$)  and freezing ($T_{\rm F}$) temperatures are respectively defined by sharp increases and decreases of the real part of the ac-susceptibility. $T_{\rm F}$ marks the onset of strong irreversibilities observed in the SG and mixed M2 phases. Below $x_{\rm C}$, another transition line marks the freezing of transverse spin components and the onset of weak irreversibilities, observed in the mixed M1 phase. The "canting" temperature ($T_{\rm K}$) associated with this transition is situated between $T_{\rm C}$ and $T_{\rm F}$, as predicted by the MF model of Gabay \& Toulouse\cite{Gabay1981}. We also performed dc-magnetization measurements in the 0-5 T field range in order to verify that all samples with $x < x_{\rm C}$ retain a ferromagnetic character, while the field at which technical saturation takes place increases with $x$ under the effect of increasing magnetic frustration (Fig. \ref{fig:diagram-mag}b). From the corresponding Arrott plots (Fig. \ref{fig:diagram-mag}c), we also deduced the $x$-dependence of the spontaneous moment $M_{\rm 0} = M(\mu_{0}H_{\rm int} \rightarrow 0)$. As shown in the inset of Fig. \ref{fig:diagram-mag}c, this leads to an extrapolated zero moment for $x = 0.38(3)$, which is consistent with the literature value of the critical concentration $x_{\rm C}$ where long-range FM order is lost. 

\section*{Small-angle neutron scattering (SANS)}

%In order to study the microscopic magnetic structure of our series of a-FeMn samples, 
We have carried out a small-angle neutron scattering (SANS) experiment at the PAXY beamline (Orph\'ee Reactor, Saclay, France). A horizontal magnetic field $\mu_{0}\mathbf{H}$ up to 4 T was applied transverse to the beam direction ({\it i.e.}, in the detector plane). All data were obtained in the zero field cooled (ZFC) state at $T$ = 3 K. This temperature was chosen because {\it(i)} it is well-below $T_{\rm C}$ and $T_{\rm F}$ for all samples (Fig. \ref{fig:diagram-mag}), and {\it(ii)} allows neglecting the contribution of phonons and magnetic excitations (spin waves) to the SANS patterns. 
%Complementary measurements were performed on the spectrometer D33 of the Institut Laue Langevin in the field cooled (FC) state to check the influence of the cooling field (see Methods for details). 
Data were corrected and calibrated as described in the Supplementary Information.   

\begin{figure}[!ht]
	\includegraphics[width=\textwidth]{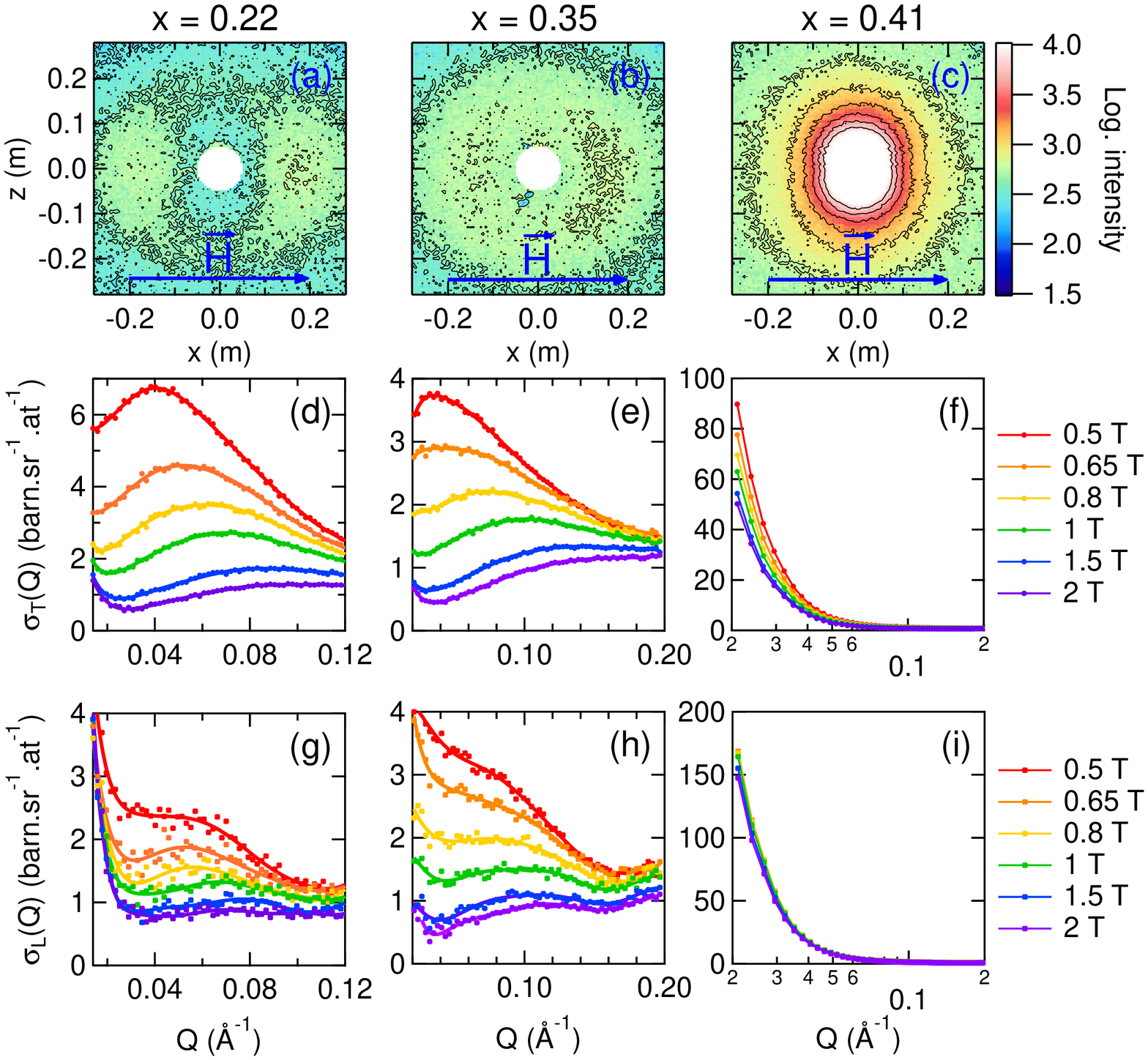}
	\caption{{\bf SANS data on a-FeMn in the RSG and SG phases --} {\bf (a-c)} Scattering maps recorded at T = 5 K under an applied magnetic field of 1.5 T. {\bf (d-f)} Field-dependence of the transverse magnetic cross section $\sigma_{\rm T}(Q)$. {\bf (g-i)} Field-dependence of the longitudinal magnetic cross section $\sigma_{\rm L}(Q)$. In panels {\bf (d-i)}, we indicate the values of the applied magnetic field.}
	\label{fig:sq_sans}
\end{figure} 

Typical SANS patterns are shown in Fig. \ref{fig:sq_sans}, for compositions $x$ respectively below (Fig. \ref{fig:sq_sans}a,b) and above (Fig. \ref{fig:sq_sans}c) $x_{\rm C}$. One can immediately note differences between these two cases. On the one hand, maxima of intensity appear parallel to the applied field at a finite value of the momentum transfer $Q$ when $x < x_{\rm C}$. On the other hand, SANS from the $x = 0.41$ sample is typical of field-induced ferromagnetic-like correlations centered at $Q = 0$. In order to separate contributions of magnetic moments transverse ($T$) and longitudinal ($L$) to the applied field, we make use of the neutron selection rule which states that only components perpendicular to the scattering vector $\mathbf{Q}$ contribute to the observed scattering cross section $\sigma(\mathbf{Q})$. This translates into the following relations

\begin{equation}\label{eq:sqtl}
	\sigma_{\rm T}\left(\mathbf{Q}\right) + \sigma_{\rm bg}\left(\mathbf{Q}\right) /2 = \tilde{\sigma}\left(\mathbf{Q}\parallel\mathbf{H}\right) /2 \quad \text{and} \quad \sigma_{\rm L}\left(\mathbf{Q}\right)  + \sigma_{\rm bg}\left(\mathbf{Q}\right) /2 = \tilde{\sigma}\left(\mathbf{Q}\perp\mathbf{H}\right) -\tilde{\sigma}\left(\mathbf{Q}\parallel\mathbf{H}\right) /2 \quad ,
\end{equation}

where $\sigma_{\rm bg}$ denote the background contributions from the sample ({\it e.g.}, nuclear scattering) and $\tilde{\sigma}$ the full observed scattering within sectors of 60$^{\circ}$ opening angle, parallel or perpendicular to $\mathbf{H}$. Therefore, a radial integration of the SANS data along the horizontal and vertical direction allows retrieving the $Q$-dependences of $\sigma_{\rm T}$ and $\sigma_{\rm L}$ independently, assuming an isotropic $\sigma_{\rm bg}$. 

The result of such procedure is shown in Figs. \ref{fig:sq_sans}d-i. In the weakly frustrated $x = 0.22$ RSG sample, the intensity is clearly enhanced along the field direction, i.e. for $\mathbf{Q}\parallel\mathbf{H}$, showing that the contribution of spin components transverse to the magnetic field are dominant in the explored $Q$-range, whereas the opposite behavior is observed in the $x = 0.41$ SG sample. As a general feature, we observe field-induced peaks in $\sigma_{\rm T}(Q)$ at $Q = Q_{\rm max}$ for all compositions $x < x_{\rm C}$. $Q_{\rm max}$ moves to higher values when the field increases at constant $x$, and also shows a systematic stiffening as $x$ increases towards $x_{\rm C}$. $\sigma_{\rm L}(Q)$ shows a broad maximum, but it is more difficult to point because its intensity is much smaller. 

In what follows, we focus on the transverse cross section $\sigma_{\rm T}(Q)$, which is defined in the most general case as

\begin{equation}\label{eq:sqt_interf}
	\sigma_{\rm T}\left(\mathbf{Q}\right) \propto \langle F_{\rm T}^{2}(Q) \rangle  - \langle F_{\rm T}(Q) \rangle^{2} \left[1-S_{\rm int}(Q)\right] \quad ,
\end{equation}

where $F_{\rm T}(Q)$ is the form factor of the transverse defects and $S_{\rm int}(Q)$ is an interference function that expresses the local correlations between two such defects. Assuming that the latter are organized in a liquid-like order ($S_{\rm int}(Q) \rightarrow 1$) and noting that the form factor of a "regular" vortex is null ($\langle F_{\rm T}(Q) \rangle = 0$)\cite{Mirebeau2018}, $\sigma_{\rm T}(Q)$ is simply proportional to $\langle F_{\rm T}^{2}(Q)\rangle$. 

\begin{figure}[!ht]
	\includegraphics[width=\textwidth]{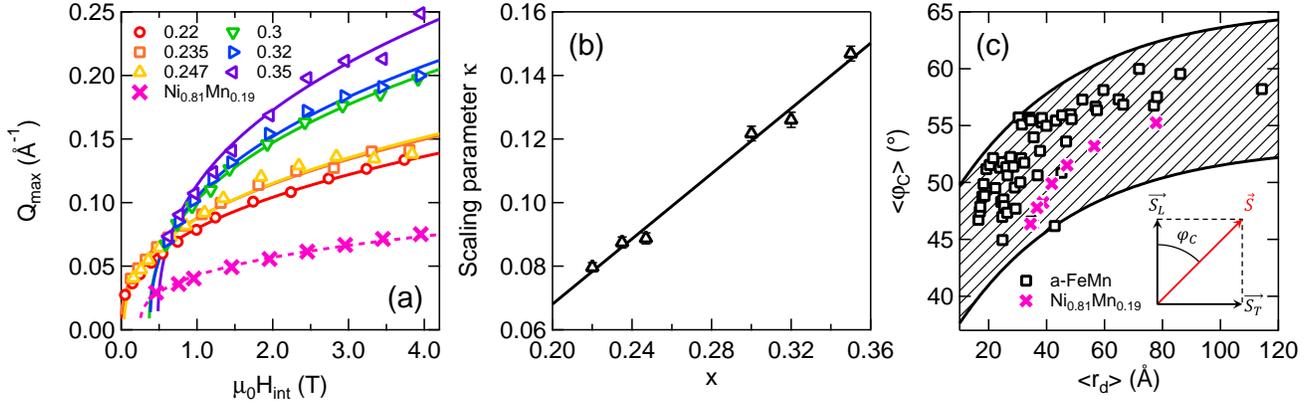}
	\caption{{\bf Scaling of vortex properties --} {\bf (a)} Field-dependence of the position $Q_{\rm max}$ of the maximum in $\sigma_{\rm T}(Q)$ for each studied compositions. Solid lines are results of a global fit of Eq. \ref{eq:scaling_law} to the data. {\bf (b)} $x$-dependence of the scaling parameter $\kappa$, extracted from a fit of Eq. \ref{eq:scaling_law} to the data of panel \textbf{(a)}. {\bf (c)} Average canting angle $\langle \varphi_{\rm C} \rangle$ as a function of the average vortex radius $\langle r_{\rm d} \rangle$. In panel \textbf{(a)} and \textbf{(c)}, data for Ni$_{0.81}$Mn$_{0.19}$ is shown for comparison (pink crosses)\cite{Mirebeau2018}. Dashed pink line is the result of a fit of Eq. \ref{eq:scaling_law} to the data.}
	\label{fig:scaling_sans}
\end{figure}

In the whole $x < x_{\rm C}$-range, the field-dependence of $Q_{\rm max}$ obeys a scaling law of the form

\begin{equation}\label{eq:scaling_law}
	Q_{\rm max} (\mu_{\rm 0}\,H_{\rm int},x) = \kappa(x) \, \left[\mu_{\rm 0}\,\left(H_{\rm int} - H_{\rm 0}(x)\right)\right]^\gamma \quad ,
\end{equation}

where $H_{\rm 0}(x)$ is a composition-dependent saturation field, increasing with $x$. A global fit of Eq. \ref{eq:scaling_law} to the data yields a unique exponent $\gamma = 0.39(1)$ and $x$-dependent scaling parameters $\kappa(x)$ (Figs. \ref{fig:scaling_sans}a,b). Results previously obtained on Ni$_{0.81}$Mn$_{0.19}$ are also reported in Fig. \ref{fig:scaling_sans}a. In this case, a fit of Eq. \ref{eq:scaling_law} to the data yields an exponent $\gamma = 0.34(2)$, quite close to the value derived for the a-FeMn series.

In order to account for this scaling law, we propose a simple picture and provide a physical meaning for its parameters. First, we note band structure calculations of dilute FeMn alloys\cite{Campbell1967,Mirzoev2006,Schneider2018} where Mn-Mn NN interactions are AFM whereas Fe-Fe and and Fe-Mn NN interactions are FM. Based on previous SANS results on a crystalline Ni$_{0.81}$Mn$_{0.19}$ sample, as well as toy models and MC simulations \cite{Mirebeau2018}, we interpret  $\sigma_{T}(Q)$ by assuming uncorrelated defects akin to vortices, nucleated around Mn-Mn first neighbor pairs. In the simplest picture, the spins components $M_{\rm T}$ are ferromagnetically correlated and rotate over an average vortex radius $\langle r_d \rangle = \pi/Q_{\rm max}$ to compensate the transverse magnetization inside the vortex\cite{Mirebeau2018}.  
 
Using this picture, we can readily interpret the evolution of the SANS patterns with magnetic field and Mn concentration $x$. At a given $x$, $\langle r_d \rangle$ decreases with increasing the magnetic field (Fig. \ref{fig:scaling_sans}a), as the vortices progressively align along the field, albeit not necessarily in a uniform fashion. Their gradual suppression yields a steady increase of the magnetization (Fig. \ref{fig:diagram-mag}b). At a given field, $\langle r_d \rangle$ also decreases with an increase of $x$, which governs the concentration of AFM NN pairs within the samples. As we will show below, these features translate the collapse of the average exchange interaction $\langle J \rangle$.

Altogether, such scaling law suggests that the characteristic size of the defect is governed by the ratio between the magnetic field and the average exchange interaction.
To check this picture in more details, we have searched for a common law governing the bulk magnetization curves $M(H)$ at 5 K in the a-FeMn system. From the experimental curves, we find a $M \approx \left(\mu_{0}\,\left[H_{\rm int}-H_{\rm 0}\right]\right)^{1/3}$-dependence above a threshold field value $H_{\rm 0}$, which scales with the saturation field deduced from the magnetization curves (see Supplementary Information with Supplementary Fig. 3). We can compare this dependence with that observed in ferromagnets close to saturation, where the magnetic field suppresses microstructural defects. Here the magnetostatic exchange length $\Lambda$ which controls the defect size is defined as\cite{Weissmuller1999}

\begin{equation}\label{eq:l_exchange}
	\Lambda = \left(\frac{2A}{\mu_0 M^2}\right)^{1/2} \quad ,
\end{equation}
where $A$ is the exchange stiffness and $M$ the bulk magnetization. Identifying $A$ with the average exchange term $\langle J \rangle$ and $\Lambda$ with $\langle r_{\rm d} \rangle$ leads to the following dependence  for $Q_{\rm max}$

\begin{equation}\label{eq:qmax_theo}
	Q_{\rm max} \approx \frac{\left(\mu_{0}\,\left[H_{\rm int}-H_{\rm 0}\right]\right)^{1/3}}{\langle J \rangle^{1/2}} \quad .
\end{equation}

 which is quite close to the dependence found experimentally (Eq. \ref{eq:scaling_law}), noticing that the  experimental value of the exponent $\gamma =0.39 (1)$ is slightly above the value $1/3$ from macroscopic magnetization. This comparison however confirms that the average exchange interaction and the applied magnetic field are the main ingredients needed to control the behaviour of the observed magnetic defects, although additional anisotropic exchange terms (such as the Dzyaloshinskii-Moriya interaction) could play a minor role. A perfect mapping of the two cases is in fact not expected, especially for the strongly frustrated RSGs, where the magnetization does not show any clear saturation plateau. Taking this analysis into account, one can however tentatively evaluate the average exchange constant from the scaling parameter $\kappa$ by $\langle J \rangle =\kappa^{-1/\gamma_{\rm eff}}$, with $1/3 \leq \gamma_{\rm eff} \leq 0.39$. 

We can further define the canting angle $\langle \varphi_{\rm C} \rangle$, averaged over the vortex size, by the expression

\begin{equation}\label{eq:theta_canting}
	\langle \varphi_{\rm C} \rangle = \frac{\text{arctan}\left(\langle M_{\rm T}^2 \rangle / \langle M_{\rm L}^2 \rangle\right)}{\langle r_{\rm d} \rangle} \approx \text{arctan}\left(\frac{\sigma_{\rm T}(q_{\rm max})}{\sigma_{\rm L}(q_{\rm max})}\right) \cdot q_{\rm max} \quad .
\end{equation}

The result is shown in Fig. \ref{fig:scaling_sans}c, which illustrates the correlation between the vortex average radius and $\langle \varphi_{\rm C} \rangle$. $\langle \varphi_{\rm C} \rangle$ is maximum at low fields when the vortex size is the largest (around 120 \AA), and reaches values of 55-60 degs. As for comparison, the canting angle deduced from the M\"ossbauer measurements of the $^{57}$Fe hyperfine field in the $x$ = 0.235 sample, is around 35(7) degs, with a large distribution\cite{Mirebeau1986}.  As the field increases, the vortex size decreases and the canting angles decreases as well, reaching values around 45 degrees at the smallest vortex size of $\approx$ 20 \AA. To summarize this point, the average canting angle  $\langle \varphi_{\rm C} \rangle$ increases as the transverse spin components and magnetic disorder in the transverse plane increase. Corresponding data, extracted from our previous SANS experiment on Ni$_{0.81}$Mn$_{0.19}$ fall within the same range and follows a very similar trend\cite{Mirebeau2018}. This suggests that the observed vortices are relatively independent on the sample nature (single crystalline or amorphous) and could represent an immanent feature of the large family of RSG. 

\section*{Monte Carlo simulations}

In order to get a deeper insight onto the properties of the magnetic defects evidenced in our SANS experiments, we have carried out a series of Monte Carlo simulations on 2d square lattices containing 10$^4$ spins. The model is described by the following classical Hamiltonian:

\begin{equation}\label{eq:hamiltonian_MCsimuls}
	\mathcal{H} = - \sum_{ij} J_{ij} \, \mathbf{S}_{i} \cdot \mathbf{S}_{j} - \alpha H \sum_{i} S_{i}^{z} \quad ,
\end{equation}

where $\mathbf{S}_{i,j}$ are Heisenberg spins with $|\mathbf{S}_{i,j}| = 1$, $J_{ij}$ are random independent variables taking the value $\pm 1$, $\alpha = \mu_{\rm B} / k_{\rm B} \approx 0.672$ is a coupling constant and the magnetic field $H$ is applied along the $z$ direction. The first sum in Eq. \ref{eq:hamiltonian_MCsimuls} runs over NN pairs. All simulations started by generating random spin configurations at a temperature $T = 2J$, where a concentration $x$ of "impurities" (i.e., analogs of Mn ions) are scattered within an otherwise ferromagnetic matrix (i.e., analogs of Fe ions). The following rule is then applied to calculate the sign of the nearest-neighbors exchange terms: two nearest-neighbor impurities will be coupled antiferromagnetically ($J_{ij} = -1$) while all other pairs will be coupled ferromagnetically ($J_{ij} = 1$). This situation is meant to map quite closely the one expected from band structure calculations\cite{Campbell1967,Mirzoev2006,Schneider2018}, let alone the actual atomic connectivities. The key quantity describing the MC sample is therefore the concentration $c_{\rm AFM}$ of AF bonds. In order to stick even more to the experimental situation, the system is slowly cooled down to $T = 0.01 J$ at $H = 0$ and the field is further raised in steps $\Delta H = 0.01 J$. We shall show in the following that such a simple scheme allows for a "zero-order" simulation of the properties of the RSG, and a reasonable description of the experimental observations reported in this paper.

\begin{figure}[ht]
	\begin{center}
	\includegraphics[width=\textwidth]{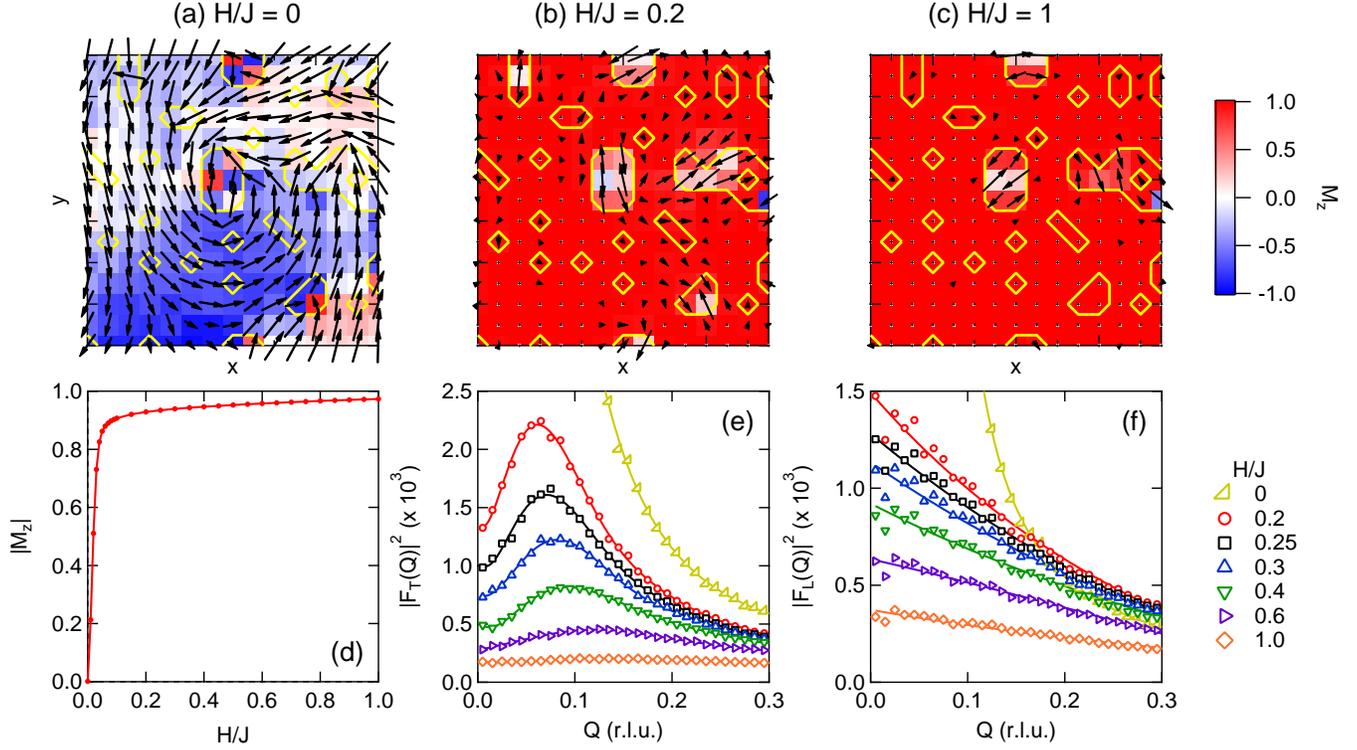}
	\end{center}
	\caption{{\bf Spin configurations and Fourier analysis of the MC simulations --} {\bf (a-c)} Snapshot of a spin configuration ($15 \times 15$ spins) obtained on a $x = 0.23$ sample ($c_{\rm AFM} \approx 0.05$) at $T/J = 0.01$ for \textbf{(a)} $H/J = 0$, \textbf{(b)} $H/J = 0.2$ (slightly above technical saturation) and \textbf{(c)} $H/J = 1$. Arrows represent the in-plane magnetization while color map shows the out-of-plane ({\it i.e.}, parallel to the applied magnetic field) component $M_{\rm Z}$. Yellow lines border the regions where impurity spins are located, anchoring the field-induced localized vortex-like textures (see text). {\bf (d)} Field-dependence of the average longitudinal magnetization $|M_{\rm Z}|$. {\bf (e-f)} Squared Fourier transforms of the transverse \textbf{(e)} and longitudinal \textbf{(f)} spin correlations for different values of $H/J$.}
	\label{fig:MC}
\end{figure}

We now focus on the case of a weakly frustrated RSG sample (concentration of AFM bonds $c_{\rm AFM} \approx 0.05$) to investigate the spin configurations  and the corresponding Fourier maps as the magnetic field increases (additional cases, displaying essentially similar behaviors, are addressed in the Supplementary Information with Supplementary Figs. 9 and 10). In the zero-field or low field region, vortex-like structures are observed around AFM NN pairs, coexisting with domains walls of large length scale which separate the magnetic domains. As shown in Fig. \ref{fig:MC}a, these domain walls can involve transverse chiral components as well as local defects. However, a Fourier transform of the spin maps show that they \textit{do not} yield a maximum of the scattered intensity as for the vortex-like structures, but rather a huge increase of the intensity at low $Q$ values. As the field increases, these walls are rapidly suppressed, leading to a strong increase of the magnetization, and to the observation of isolated vortices. Such textures are nucleated randomly in the sample around AFM NN pairs (Fig. \ref{fig:MC}a-c), so that in zero (or small) applied field, they could form both in the ferromagnetic domains and in the domain walls (Fig. \ref{fig:MC}a). However, they are observed in the Fourier maps (i.e. by SANS) only when the field is high enough to suppress the contribution of the domain walls. In other words, vortices emerge from a ferromagnetic \textit{vacuum}. In this high field region, magnetization shows a quasi-plateau (Fig. \ref{fig:MC}d) while the vortex contributions remains alone, leading to a clear maximum of the calculated scattered intensity at a finite $Q$-value. The vortices shrink as the field increases further, and they slowly disappear together with the AFM pairs which nucleate them. The complete  destruction of all AFM pairs should only occur at very high fields, much larger than the exchange interaction ($H \gg J$ ). Fourier transforms of the spin maps in the region of the magnetization plateau shows features very similar to the experimental ones, both for the transverse and longitudinal contributions to the cross section (compare Figs. \ref{fig:MC}e,f and Figs. \ref{fig:sq_sans}d,e,g,h). Finally, we note that while being resilient to very large applied fields, the vortex-like textures obtained in the simulations have a vanishing topological charge, most likely due to the ill-defined FM vacuum endowing them with irregular shapes. 

\section*{Discussion}

As our key experimental result, we have shown that in a frustrated ferromagnetic system, vortex-like defects are a characteristic feature of the RSG  ground state. Their average size decreases with decreasing the average exchange interaction, and the vortices disappear in the true SG. The average vortex size $\langle r_{\rm d} \rangle$ also decreases with increasing the applied magnetic field. The whole behaviour is captured by a scaling law governing $\langle r_{\rm d} \rangle$, where the only ingredient is the ratio of the internal magnetic field to the average exchange interaction. Similar laws derived from magnetostatics govern the field behaviour of different macroscopic quantities. For example, one can quote the quasi saturated magnetization of ferromagnets with microstructral defects, the magnetization of type II superconductors, or the thickness of Bloch walls in ferromagnets.
 
In the RSGs, the presence of vortex-like defects up to the critical concentration, and their collapse in the SG phase when the average exchange interaction becomes smaller than the width of its distribution, strongly supports the existence of a critical line between  RSG and SG regions. Our observations therefore support a MF description of the RSG phase diagram, rather than the crossover evolution towards FM breakdown predicted by random field arguments. We however recall that the original MF model of Gabay-Toulouse \cite{Gabay1981}, although being able to correctly describe the experimental $(x,T)$-phase diagram, cannot predict any defect, since the transverse spin component is randomly distributed in the transverse plane. The present observation can therefore help refining the current models for the RSG problem, by considering the observed magnetic microstructure. 
  
Altogether, our SANS results combined with MC simulations suggest two complementary phenomena: \textit{(i)} the vortices emerge from an average ferromagnetic medium acting as a vacuum field, required for their stabilization and \textit{(ii)} they protect the ferromagnetic domains from breaking down under the influence of magnetic fustration. The MC calculations strongly suggest that the vortices are nucleated around AFM NN pairs. In real samples, these pairs likely arise from Mn-Mn first neighbours, as suggested by band structure calculations in Fe-Mn\cite{Campbell1967,Mirzoev2006,Schneider2018}.  
  
In the weakly frustrated RSG, vortices and domain walls can be clearly distinguished by combining magnetization, SANS and MC simulations. The domain walls recall those observed in non-frustrated ferromagnets, but they involve AFM bonds, which induce magnetic defects where the spin canting is locally enhanced, and which act as pinning centers at low temperature. Below $T_{\rm F}$, this process leads to a strong decrease of the susceptibility $\chi(T)$ and to strong irreversibilities of the magnetization $M(H)$. This picture  is supported by electron microscopy\cite{Senoussi1988}, neutron depolarisation\cite{Mirebeau1990} and recent acoustic absorption measurements\cite{Kustov2017}, which clearly show that LRMO and $\mu$m-sized domains are preserved in the ground state of the weakly frustrated RSG. As the field increases, the domain walls are washed out by low fields, whereas vortices persist up to much higher fields, where their contribution to the SANS can be clearly identified. In the highly frustrated RSG, the distinction between the vortices and the domain walls smears; the average domain size decreases and becomes comparable to the domain wall thickness, and the magnetization plateau dissappears. The vortex contribution is still clearly observed in the SANS data, in sharp contrast with the SG sample. 
 
MC simulations are in turn crucial to refine the above picture, and already extend our results to a field range (or more precisely $H/J$ range) inaccessible to experiment. The good agreement between the Fourier transform of the MC spin maps and our experimental results should be noticed, considering that the simulated case is over-simplified with respect to the experimental one. We outline here that the MC spin maps show a huge amount of disorder around the local defects which nucleate the vortices. Considering the chemical disorder, many different types of vortices could \textit{a priori} exist in the sample and they are indeed observed in the MC spin maps. However, their average size can be determined without ambiguity, as it leads to a maximum in $Q$-space, the position $Q_{\rm max}$ of which is tuned by the $H/J$ ratio. 
  
To conclude, we briefly compare the above vortices with the topological defects observed in ferromagnets submitted to weak random fields. Quite generally, topological defects are expected when the number of spin components $n$ is such that $n \leq d+1$ where $d$ is the dimension of space\cite{Toulouse1976}, namely in all experimental cases. For instance, non-singular skyrmions with a finite topological charge are observed in the ($n = 3$, $d = 2$) case\cite{Proctor2014}. Their existence is a consequence of crossing points between lines where all RF field components cancel at the same time. This leads to the very interesting concept of "skyrmion glass", composed of regions with oscillating positive and negative topological charges, and sizes scaling that of the IM domains\cite{Chudnovsky2018}. Importantly, these defects, which prevent the magnetization from collapsing, should lead to a measurable topological Hall effect (THE). Conversely, the vortices stabilized by magnetic frustration (induced by competing interactions and bare interaction randomness) have a very small topological charge due to their very irregular shape, but they could also yield a peculiar Hall signal. In this context, it is worth noting that an anomalous Hall effect was actually predicted\cite{Kawamura2003} and observed in AuFe RSG or SG alloys\cite{Pureur2004,Fabris2006}, as a probe of non coplanar (chiral) spin configurations. A quantitative study of the field-dependent Hall response of a-FeMn above and below $x_{\rm C}$ could refine the description of the RSG ground state, given that the defects involved in the two regimes likely have different natures.
%have very different topologies.

\section*{Acknowledgements}
We thank S. Gautrot (LLB) and M. Bonnaud (ILL) for their assistance during the small-angle neutron scattering experiments at the PAXY and D33 instruments, respectively.

\section*{Materials and methods}

\noindent {\bf Materials.} The amorphous samples of (Fe$_{1-x}$Mn$_{x}$)$_{75}$P$_{16}$B$_{3}$Al$_{3}$ (0.22 $\leq$ x $\leq$ 0.41) used in this study were prepared using the "wheelbarrow" technique, which consists in casting molten alloy with the desired composition on a spinning wheel. Being a strong neutron absorber, $^{10}$B was replaced with isotopic $^{11}$B. Samples were cut in foils of about 1 cm$^2$ surface with thicknesses varying from 30-70 $\mu$m. These foils were piled up in order to increase the total sample thickness and yield a large enough sample mass for the small-angle neutron scattering experiments. Conversely, individual foils were cut into rectangular pieces, having a height to width ratio close to 2, for the magnetic measurements.

\noindent {\bf Magnetic measurements.} The ac-susceptibility of the a-FeMn samples have been obtained using a Quantum Design Physical Properties Measurement System (PPMS, Dynacool 9 T, Laboratoire L\'eon Brillouin, France). Magnetization curves were measured using a Quantum Design Superconducting Quantum Interference Device magnetometer (SQUID, MPMS-XL 5 T, Technische Universiteit Delft, The Netherlands).

\noindent {\bf Small-angle neutron scattering (SANS).} SANS experiments were performed on the PAXY instrument at the Orph\'ee reactor (Laboratoire L\'eon Brillouin, Gif-sur-Yvette, France), operated in a standard pinhole geometry. Neutron wavelength was set to 4 and 6 \AA, while keeping the sample-to-detector distance to 2.8 m. An horizontal magnetic field was applied using a cryomagnet (Oxford SM4000), allowing to reach fields of 10 T while cooling the sample down to 2 K. Additionnal SANS measurements were performed on the D33 instrument (Institut Laue Langevin, Grenoble, France)\cite{Mirebeau2015}, as described in the Supplementary Information and Supplementary Fig. 7.

\noindent {\bf Monte Carlo simulations.} Monte Carlo simulations were carried out using the "adaptative" algorithm described by Alzate-Cardona {\it et al.}\cite{Alzate-Cardona2019}. 40 maps, containing 10$^{4}$ spins sitting on the vertices of a square lattice were generated at high temperature. A concentration $x$ of "impurity" spins was scattered across the matrix in order to introduce AFM couplings within the FM matrix (all couplings had a magnitude $J$). Each sample was cooled down to $T = 0.01 J$ in zero-applied field, and the field was then increased in small steps to study the evolution of the spin configurations (Figs. \ref{fig:MC}a-c) and their Fourier transforms (Figs. \ref{fig:MC}e-f).

\pagebreak
\appendix

\begin{center}
{\bf \Large Supplementary Information}
\end{center}

In this supplement, we provide information concerning the synthesis and structural characterization of the (Fe$_{1-x}$Mn$_{x}$)$_{75}$P$_{16}$B$_{6}$Al$_{3}$ samples used in this study (Sec. \ref{sec:samples}). Composition-dependence of the magnetization and AC susceptibility data is presented in Sec. \ref{sec:magnetic}. In Sec. \ref{sec:sans}, we describe the strategy used to scale and analyze the small-angle neutron scattering (SANS) data. Finally, the details of our Monte Carlo simulations on systems with varying antiferromagnetic bond concentration are given in Sec. \ref{sec:mcsims}.

%%%%%%%%%%%%%%%%%%%%%%%%%%%%%%
%%%%%%%%%%%%%%%%%%%%%%%%%%%%%%
%%%%%%%%%%%%%%%%%%%%%%%%%%%%%%
\section{Samples synthesis and structural characterization}
\label{sec:samples}
%%%%%%%%%%%%%%%%%%%%%%%%%%%%%%
%%%%%%%%%%%%%%%%%%%%%%%%%%%%%%
%%%%%%%%%%%%%%%%%%%%%%%%%%%%%%

Amorphous samples of (Fe$_{1-x}$Mn$_{x}$)$_{75}$P$_{16}$B$_{6}$A$_{3}$ (0.22 $\leq x \leq$ 0.41), herafter named "a-Fe$_{1-x}$Mn$_{x}$", were prepared using the "wheelbarrow" technique, which consists in casting molten alloy with the desired composition on a spinning wheel, by J. Bigot (Centre d'\'Etudes de Chimie—Metallurgie, Vitry sur Seine). Being a strong neutron absorber, $^{10}$B ($\sigma_{\rm abs} = 3835$~barn) was replaced with isotopic $^{11}$B ($\sigma_{\rm abs} = 0.0055$~barn). Samples were cut in foils of about 1 cm$^{2}$ surface with thicknesses varying from 30-70 $\mu$m. These foils were then piled up in order to increase the total sample thickness and yield a large enough sample mass for the small-angle neutron scattering (SANS) experiments (see Sec. \ref{sec:sans}).\\

\noindent{\bf Density --} The density $d_{\rm a-Fe_{\rm 1-x}Mn_{\rm x}}$ of these materials is an important value, allowing to calibrate the magnetization and SANS data. However, it is rather difficult to measure it directly, given the small thickness of the foils. It can nevertheless be estimated using the \emph{random close packing} approximation in a hard-sphere model. The maximum density is 64 \% of that of the densest crystalline arrangement, namely fcc, with a compacity of 0.74. Taking the atomic masses $m_{\rm n}$ (with $n =\,$ \{Fe, Mn, P, B, Al\}) into account, one gets: 

\begin{equation}\label{eq:calc_density}
	d_{\rm a-Fe_{\rm 1-x}Mn_{\rm x}} = \frac{0.75 \cdot \left[\left(1-x\right)\,m_{\rm Fe}+x\,m_{\rm Mn}\right] + 0.16\,m_{\rm P} + 0.06\,m_{\rm B} + 0.03\,m_{\rm Al}}{m_{\rm Fe}} \cdot \frac{0.64}{0.74} \cdot d_{\rm fcc\,Fe} \quad ,
\end{equation}

\noindent where $d_{\rm fcc\,Fe}$ = 8.879 g$\cdot$cm$^{-3}$ is the density of fcc Fe. Whenever relevant, Eq. \ref{eq:calc_density} is used to scale the data presented in the main text and this supplementary material.\\

\begin{figure}[!ht]
\begin{center}
\includegraphics[height=6cm]{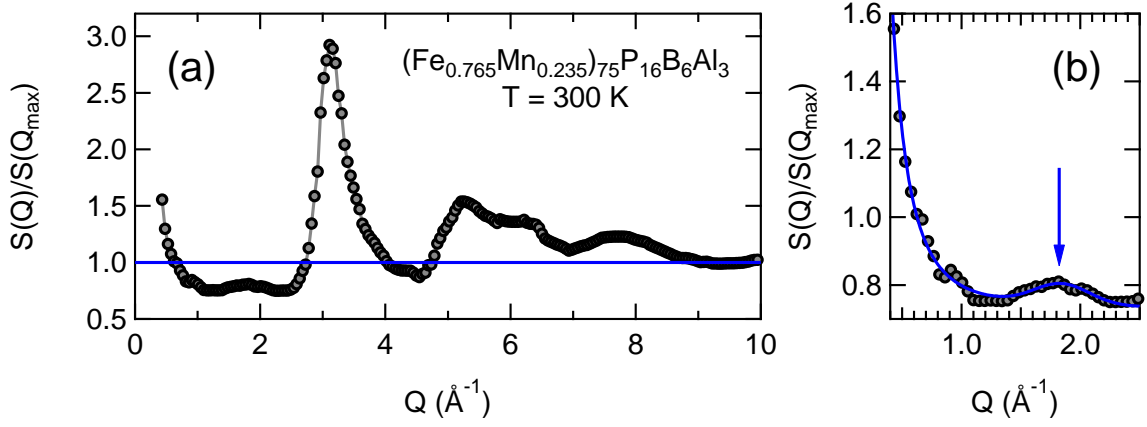}
\caption{\label{figS1} \textbf{(a)} Diffraction pattern of a-Fe$_{0.765}$Mn$_{0.235}$ (taken from Mirebeau\cite{Mirebeau1987}). Data is normalized such that $S(Q) \rightarrow 1$ in the limit of large $Q$s. \textbf{(b)} Zoom into the low $Q$ part of the diffraction pattern, showing the prepeak at $Q \approx 1.7~\text{\AA}^{-1}$. Line is a fit of a power law to the data (see text).}
\end{center}
\end{figure}

\noindent{\bf Amorphous nature of the samples --} The amorphous nature of the samples can be assessed using neutron diffraction. As shown in Fig. \ref{figS1}a in the case of a-Fe$_{0.765}$Mn$_{0.235}$, the \emph{normalized structure factor} $S(Q)/S(Q_{\rm max})$ (where $Q_{\rm max}$ is the largest momentum transfer reached in the experiment) lacks Bragg reflections and is characteristic of an amorphous (liquid-like) compound. The small-$Q$ region of the pattern (see Fig. \ref{figS1}b) is marked by an upturn, which is well-described by a power law of the form $a_{\rm p}/Q^{p}$ with an exponent $p = 2.97(12)$ (typical of surface roughness, yielding $p = 3$). The presence of a prepeak, before the main structural one, indicates a possible clustering of the main chemical specie. Its intensity is however magnified due to the fact that the scattering lengths of the main constituents (i.e., Fe and Mn) have opposite signs.

\noindent The \emph{pair distribution function} (PDF) $g(r)$ is obtained from $S(Q)/S(Q_{\rm max})$ using

\begin{equation}\label{eq:pdf}
	g(r) = 1 + \frac{1}{2\pi^{2}\,\rho_{0}\,r} \, \int_{0}^{Q_{\rm max}} Q \, \left[\frac{S(Q)}{S(Q_{\rm max})}-1\right] \, \sin \left(Qr\right) \, dQ \quad ,
\end{equation}

\noindent where $\rho_{0}$ is the atomic number density. Finally, one obtains the \emph{radial distribution function} (RDF) $\rho(r)$ from the PDF via

\begin{equation}\label{eq:rdf}
	\rho(r) = 4\pi \, \rho_{0} \, r^2 \, g(r)
\end{equation}

\begin{figure}[!ht]
\begin{center}
\includegraphics[height=6cm]{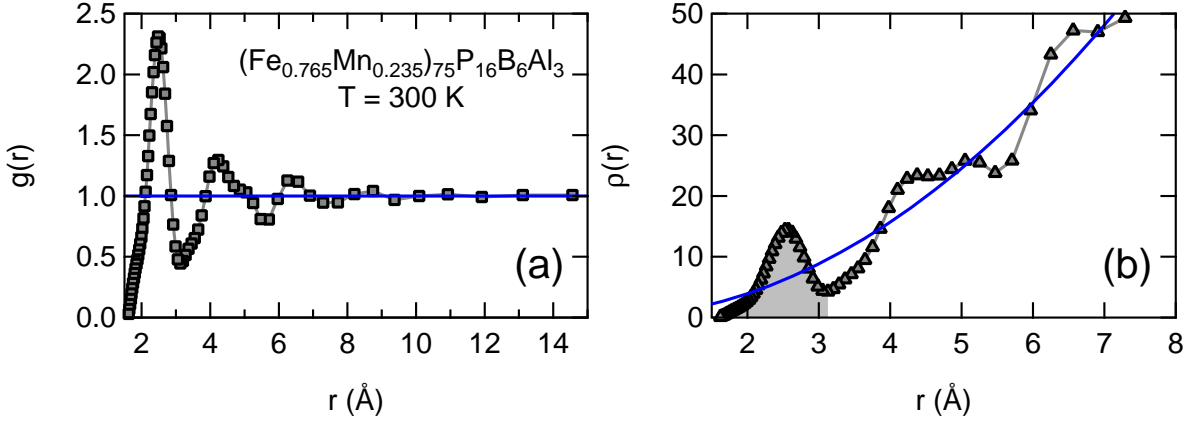}
\caption{\label{figS2}\textbf{(a)} Pair distribution function $g(r)$ and \textbf{(b)} radial distribution function $\rho(r)$ calculated from the data of Fig. \ref{figS1}.}
\end{center}
\end{figure}

\noindent This procedure allows determining the coordination numbers $z_{\rm n}$, {\it i.e.} the number of atoms in the n$^{th}$ shell surrounding any central atom, by integrating $\rho(r)$ within the $r$-range bounded by its first two minima (shaded region in Fig. \ref{figS2}b). For $n = 1$, we find $z_{1} \approx 10.6$ and $r_{1} \approx 2.6\,$\AA, in good agreement with results obtained on amorphous Fe powder~\cite{Bellissent1993} and liquid Fe~\cite{Waseda1970}

%

%%%%%%%%%%%%%%%%%%%%%%%%%%%%%%
%%%%%%%%%%%%%%%%%%%%%%%%%%%%%%
%%%%%%%%%%%%%%%%%%%%%%%%%%%%%%
\newpage
\section{Macroscopic magnetic properties}
\label{sec:magnetic}

%%%%%%%%%%%%%%%%%%%%%%%%%%%%%%
%%%%%%%%%%%%%%%%%%%%%%%%%%%%%%
%%%%%%%%%%%%%%%%%%%%%%%%%%%%%%
\subsection{Magnetization}

\begin{figure}[!ht]
\begin{center}
\includegraphics[width=\textwidth]{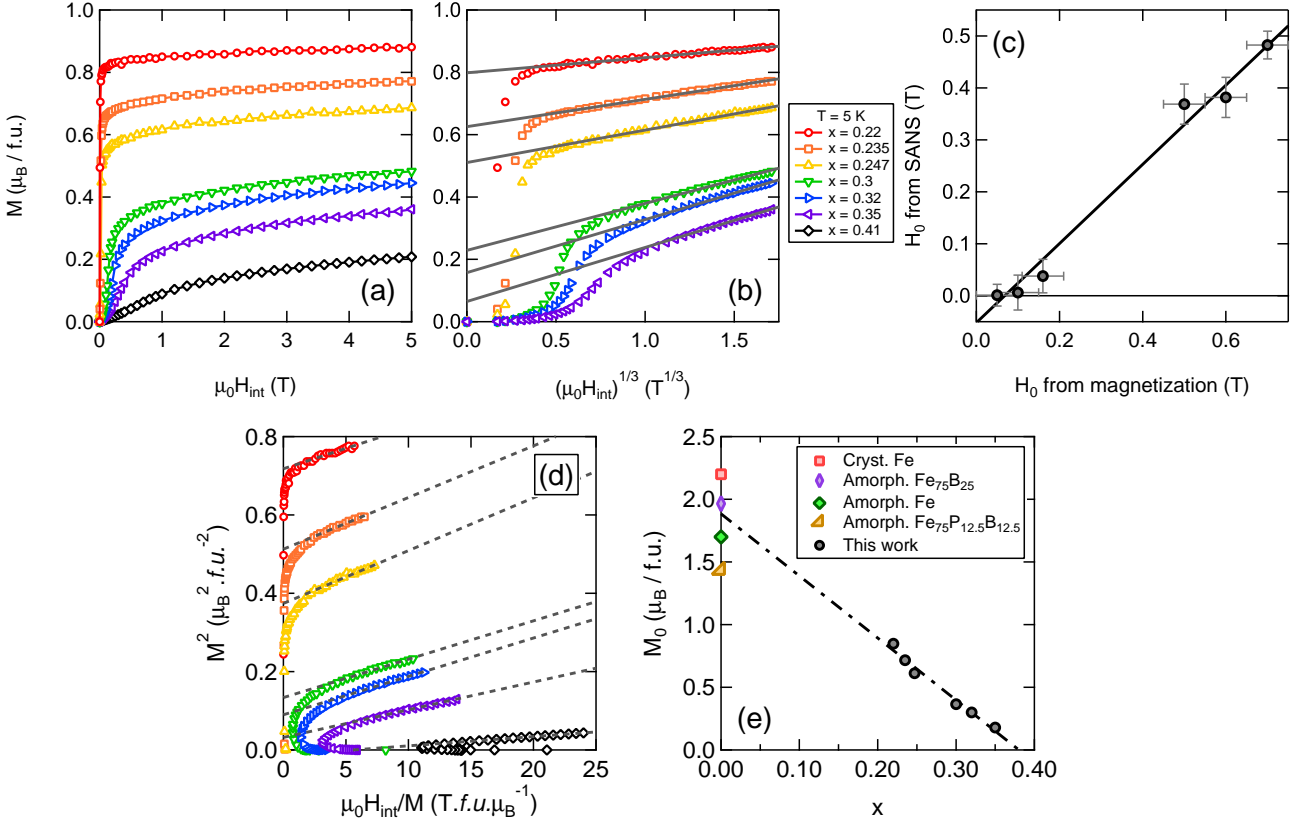}
\caption{\label{figS4}Macroscopic magnetization $M$ of the studied a-FeMn samples as a function of \textbf{(a)} $\mu_{0}H_{\rm int}$ and \textbf{(b)} $\left(\mu_{0}H_{\rm int}\right)^{1/2}$. \textbf{(c)} Saturation field value $H_{\rm 0}$ deduced from SANS data as a function of $H_{\rm 0}$ deduced from magnetization curves. \textbf{(d)} Low temperature Arrott plots (T = 5 K). \textbf{(e)} $x$-dependence of the saturated moment, inferred from the Arrott plots. It vanishes for $x \approx 0.38$ and extrapolates to a value of $\approx 1.86\,\mu_{\rm B}\cdot$f.u.$^{1}$ for $x \rightarrow 0$. This value is compared with previous results for crystalline and amorphous Fe.}
\end{center}
\end{figure}

The field-dependences of the magnetization $M$ of the a-Fe$_{1-x}$Mn$_{x}$ samples were measured using a MPMS-XL 5T Quantum Design SQUID magnetometer. The samples were zero-field cooled from $T \gg T_{\rm C}, T_{\rm F}$ down to 5 K, and specific care was taken to avoid the presence of a residual field. Subsequently, the measurements were performed by stepwise increasing the magnetic field. Samples masses of the order of several mg were used in order to be able to accurately scale the magnetic moment in Bohr magneton per formula unit ($\mu_{\rm B}$/f.u.). As shown in Fig. \ref{figS4}a, the field value at which magnetization reaches quasi-saturation ($H_{\rm 0}$) increases with increasing $x$, underscoring the increasing magnetic frustation. In all cases, $M$ retains a finite slope up to the largest fields, as a result of the gradual collapse of the vortex-like textures located around the AFM pairs (see main text). In this regime, $M$ is well-described by a law of the form $M \approx \left(\mu_{0}H_{\rm int}\right)^{1/3}$ (Fig. \ref{figS4}b). This property is used in the main text to discuss the possible scaling law governing the field-evolution of the observed nanoscopic magnetic textures. We can also define the saturation field $H_{\rm 0}$, at which $M$ acquires the $\left(\mu_{0}H_{\rm int}\right)^{1/3}$-dependence, and compare it with the $H_{\rm 0}$ extracted from the SANS scaling laws, {\it i.e.} the field above which maxima in the transverse scattering cross section $\sigma_{\rm T}$ can be defined (Fig. 3a of main text). The good correlation between these values is illustrated by Fig. \ref{figS4}c. 

The Arrott plots computed from the data of Fig. \ref{figS4}a are shown in Fig. \ref{figS4}d. All studied samples with $x < x_{\rm C}$ display a non-zero spontaneous magnetization $M_{\rm 0}$. Theses values are plotted in Fig. \ref{figS4}e. The extrapolated value of $M_{\rm 0}$ for $x \rightarrow 0$ compares well with literature values for crystalline Fe (Shull\cite{Shull1955}), and amorphous Fe$_{75}$B$_{25}$ (Cowlam \& Carr\cite{Cowlam1985}), Fe (Grinstaff \emph{et al.}\cite{Grinstaff1993}) and Fe$_{75}$P$_{12.5}$B$_{12.5}$ (Durand \& Yung\cite{Durand1977}), see Fig. \ref{figS4}.

%

%
%\begin{figure}[!ht]
%\begin{center}
%\includegraphics[width=\textwidth]{FigS5.eps}
%\caption{\label{figS5}\textbf{(a)} Low temperature Arrot plots (T = 5 K). \textbf{(b)} $x$-dependence of the saturated moment, inferred from the Arrot analysis. It vanishes for $x \approx 0.38$ and extrapolates to a value of $\approx 1.86\,\mu_{\rm B}\cdot$f.u.$^{1}$ for $x \rightarrow 0$. This value is compared with previous results for crystalline and amorphous Fe.}
%\end{center}
%\end{figure}
%

%%%%%%%%%%%%%%%%%%%%%%%%%%%%%%
%%%%%%%%%%%%%%%%%%%%%%%%%%%%%%
%%%%%%%%%%%%%%%%%%%%%%%%%%%%%%
\subsection{ac-susceptibility}

The magnetic phase diagram presented in the main text was infered through AC susceptibility measurements, performed using a Quantum Design Dynacool 9 T Physical Properties Measurement Systems (PPMS) at the Laboratoire L\'eon Brillouin. Unless otherwise stated, data presented in this section were measured under an AC field of 1 kHz frequency and 10 Oe amplitude, in zero-applied static field. In order to suppress demagnetizing field effects, we have cut the individual foils into rectangular pieces, having a height to width ratio close to 2 in each case. This however lead to very small samples masses (< 100 $\mu$g) and therefore to relatively weak signals. The AC field was applied in the sample plane, along its larger dimension. In what follows, we show how phase boundaries are deduced from maxima in $d\chi'(T)/dT$ curves.

\begin{table}[!ht]
\caption{\label{tab:ppms_samples}Dimensions of the a-Fe$_{1-x}$Mn$_{x}$ samples used for the AC susceptibility measurements.}
%\begin{ruledtabular}
\begin{center}
\begin{tabular}{ccccc}
$x$ & Height (mm) & Width (mm) \\
\hline\hline
0.22 & 4.7 & 2.3 \\
0.235 & 4.4 & 2.4 \\
0.247 & 4.2 & 2.2 \\
0.3 & 4.6 & 2.2 \\
0.32 & 5.0 & 2.2 \\
0.35 & 4.3 & 1.8 \\
0.41 & 4.0 & 2.1 \\
\end{tabular}
\end{center}
%\end{ruledtabular}
\end{table}  

\begin{figure}[!ht]
\begin{center}
\includegraphics[width=0.75\textwidth]{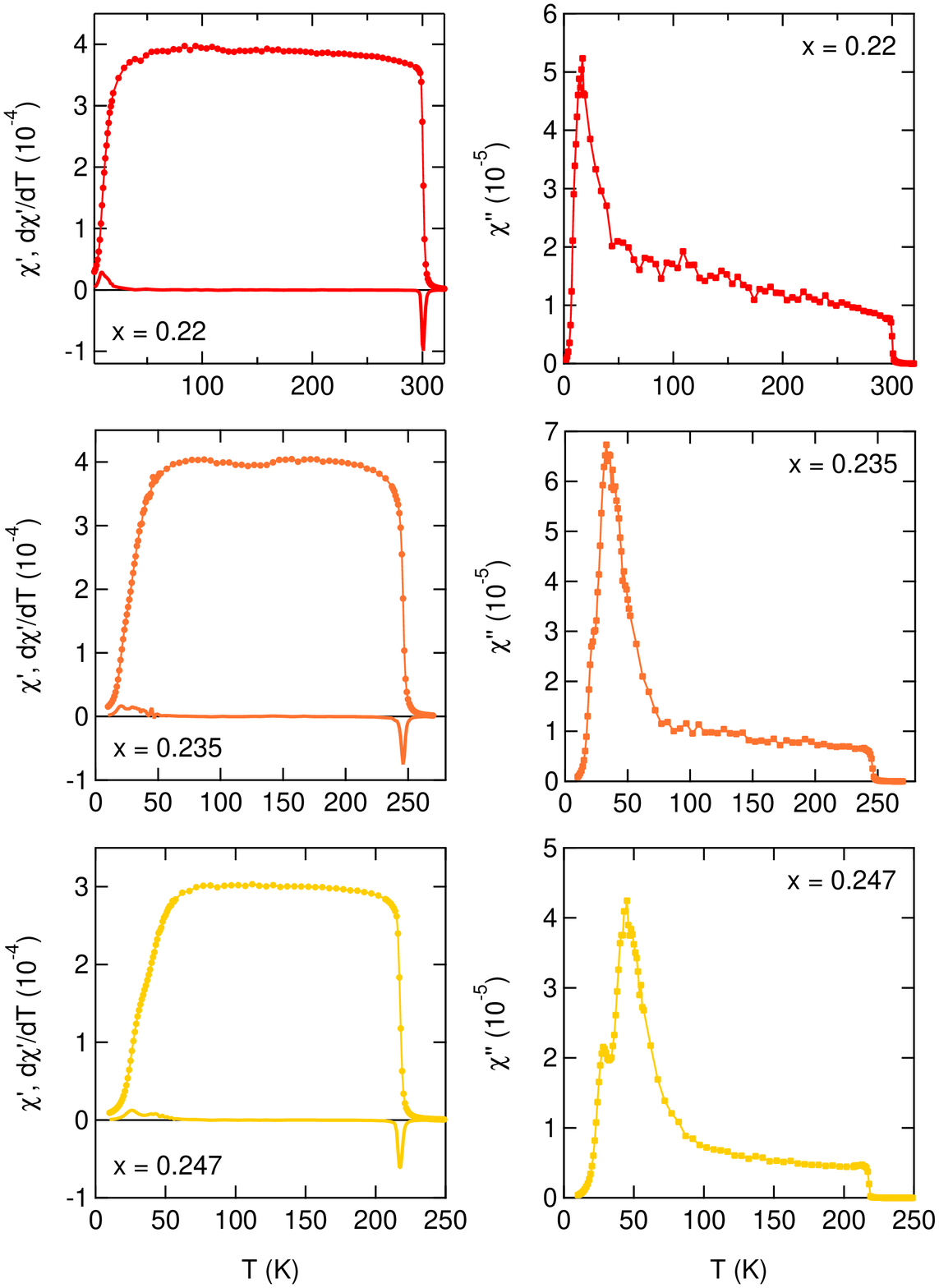}
\caption{\label{figS3a}\textbf{Zero-field AC suceptibility of a-Fe$_{1-x}$Mn$_{x}$ samples with 0.22 $\mathbf{\leq}$ x $\mathbf{\leq}$ 0.247 in zero applied field -- (Left column)} Real part of the AC susceptibility $\chi'$ (dots) and its first temperature derivative $d\chi'/dT$ (solid lines). \textbf{(Right column)} Imaginary part of the AC susceptibility $\chi''$. Data is normalized to samples' surfaces, such that $\chi'$ and $\chi''$ are expressed in emu.g$^{-1}$.Oe$^{-1}$.mm$^{-2}$, and $d\chi'/dT$ in emu.g$^{-1}$.Oe$^{-1}$.mm$^{-2}$.K$^{-1}$.}
\end{center}
\end{figure}

\begin{figure}[!ht]
\begin{center}
\includegraphics[width=0.75\textwidth]{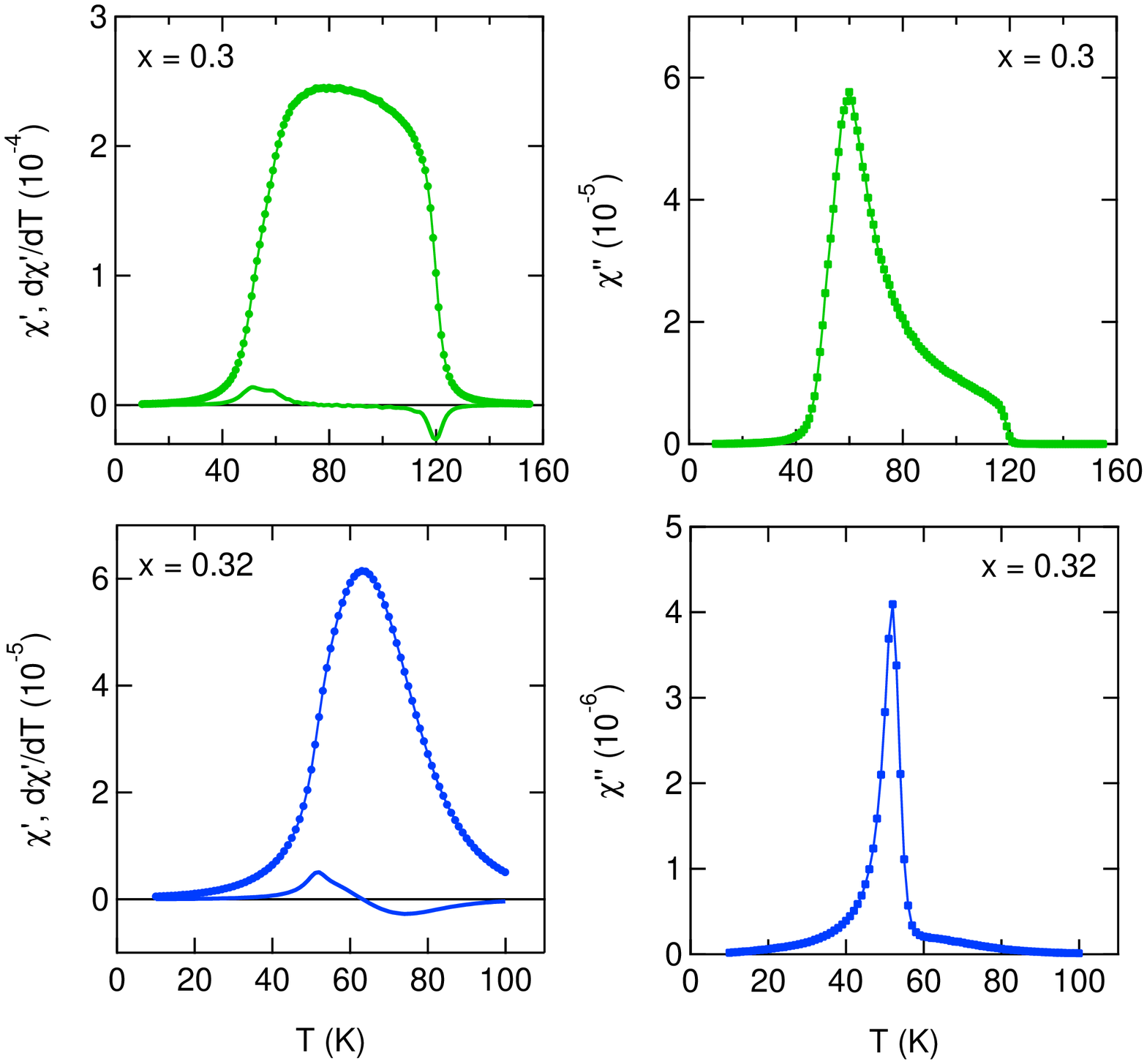}
\caption{\label{figS3b}\textbf{AC suceptibility of a-Fe$_{1-x}$Mn$_{x}$ samples with $x$ = 0.3 and 0.32 in zero applied field -- (Left column)} Real part of the AC susceptibility $\chi'$ (dots) and its first temperature derivative $d\chi'/dT$ (solid lines). \textbf{(Right column)} Imaginary part of the AC susceptibility $\chi''$. Data is normalized to samples' surfaces, such that $\chi'$ and $\chi''$ are expressed in emu.g$^{-1}$.Oe$^{-1}$.mm$^{-2}$, and $d\chi'/dT$ in emu.g$^{-1}$.Oe$^{-1}$.mm$^{-2}$.K$^{-1}$.}
\end{center}
\end{figure}

\begin{figure}[!ht]
\begin{center}
\includegraphics[width=0.75\textwidth]{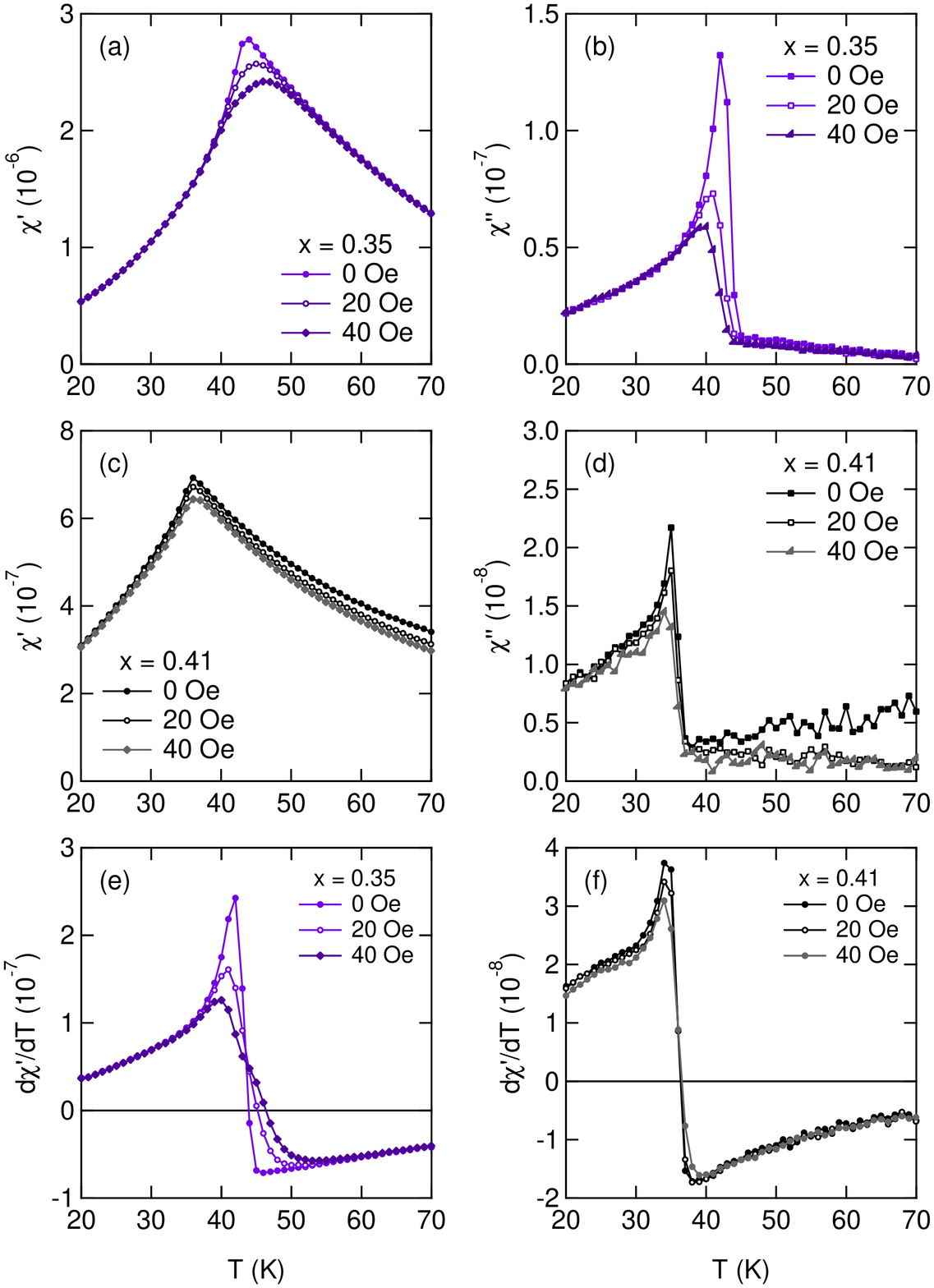}
\caption{\label{figS3c}\textbf{AC suceptibility of a-Fe$_{1-x}$Mn$_{x}$ samples with $x$ = 0.35 and 0.41 in zero and small applied field -- (a,b)} Real ($\chi'$) and imaginary ($\chi''$) part of the AC susceptibility of a-Fe$_{0.65}$Mn$_{0.35}$. \textbf{(c,d)} Real ($\chi'$) and imaginary ($\chi''$) part of the AC susceptibility of a-Fe$_{0.59}$Mn$_{0.41}$. \textbf{(e,f)} First temperature-derivative of $\chi'$. While a weak applied field has a substantial effect of the high-temperature minimum in $d\chi'/dT$ for the $x$ = 0.35 sample, it remains unchanged in the $x$ = 0.41 case. In all panels, data is normalized to samples' surfaces, such that $\chi'$ and $\chi''$ are expressed in emu.g$^{-1}$.Oe$^{-1}$.mm$^{-2}$, and $d\chi'/dT$ in emu.g$^{-1}$.Oe$^{-1}$.mm$^{-2}$.K$^{-1}$.}
\end{center}
\end{figure}

The temperature-dependence of the real ($\chi'$) and imaginary ($\chi''$) part of the AC susceptibility of a-Fe$_{1-x}$Mn$_{x}$ samples with $0.22 \leq x \leq 0.41$ is shown in Figs. \ref{figS3a}-\ref{figS3c}. For compositions $x \leq 0.32$, the Curie ($T_{\rm C}$) and spin freezing ($T_{\rm F}$) temperatures are easily evidenced by well-separated extrema in the first temperature derivative of $\chi'$ (Figs. \ref{figS3a} and \ref{figS3b}), ({\it i.e.} using the same procedure as used by Yeshurun\cite{Yeshurun1981}). The $x = 0.35$ case, located very close to the RSG-SG thershold composition $x_{\rm C} \approx 0.36$, is more difficult to anlyze. At first glance, its ac-susceptibility is very close to that of the pure SG with $x = 0.41$ (Fig. \ref{figS3c}). However, a modest field has a large impact on the $d\chi'/dT$ of the $x = 0.35$ sample as opposed to the $x = 0.41$ one. This suggests a remanence of ferromagnetism in the former, disappearing in the latter case.  

In all cases, $\chi''$ peaks at temperatures slightly higher than $T_{\rm F}$, while it falls off to $\approx 0$ around $T_{\rm C}$. Since $T_{\rm F}$ is known to depend on the ac-field frequency $f_{\rm ac}$, we have measured the temperature-dependence of the ac-susceptibility of all samples for 100 Hz $\leq f_{\rm ac} \leq$ 10 kHz (not shown). The various $T_{\rm F}$ shown in Fig. 1a of the main text are the results of extrapolations to $f_{\rm ac} = 1$\, Hz.

%

%%%%%%%%%%%%%%%%%%%%%%%%%%%%%%
%%%%%%%%%%%%%%%%%%%%%%%%%%%%%%
%%%%%%%%%%%%%%%%%%%%%%%%%%%%%%
\clearpage
\newpage
\section{Small-angle neutron scattering}
\label{sec:sans}
%%%%%%%%%%%%%%%%%%%%%%%%%%%%%%
%%%%%%%%%%%%%%%%%%%%%%%%%%%%%%
%%%%%%%%%%%%%%%%%%%%%%%%%%%%%%
\subsection{Experimental geometry}

The small-angle neutron scattering (SANS) experiment described in the main text was performed on the PAXY instrument at the Orph\'ee reactor (LLB, Gif-sur-Yvette, France). We have used a standard pinhole geometry, with parameters given in Tab. \ref{eq:paxy_params}.

\begin{table}[!ht]
\caption{\label{eq:paxy_params}Parameters used for the SANS experiment on the PAXY instrument.}
%\begin{ruledtabular}
\begin{center}
\begin{tabular}{lcl}
\hline
Neutron wavelength & $\rightarrow$ & 4 and 6 \AA \\ 
Source aperture (diameter) & $\rightarrow$ & 16 mm \\
Collimation length & $\rightarrow$ & 2.25 m \\
Sample aperture (diameter) & $\rightarrow$ & 5 mm \\
Sample-to-detector distance & $\rightarrow$ & 2.8 m \\
\hline
\end{tabular}
\end{center}
%\end{ruledtabular}
\end{table}  

\noindent A horizontal magnetic field $H = 0 - 4\,$T was supplied using an Oxford 10 T cryomagnet (Spectromag SM4000). Throughout the experiment, field was applied perpendicular to the beam direction in order to optimally resolve the azimuthal asymmetry of the magnetic scattering (see Eq. 1 of main text).
Samples were wrapped into a thin Al foil, sandwiched between two Cd slabs (to suppress background and get a well-defined sample surface with 5 mm diameter) and stuck on a Cu frame (to insure a good thermal conduction with the thermometer). 

\begin{figure}[!ht]
\begin{center}
\includegraphics[height=8cm]{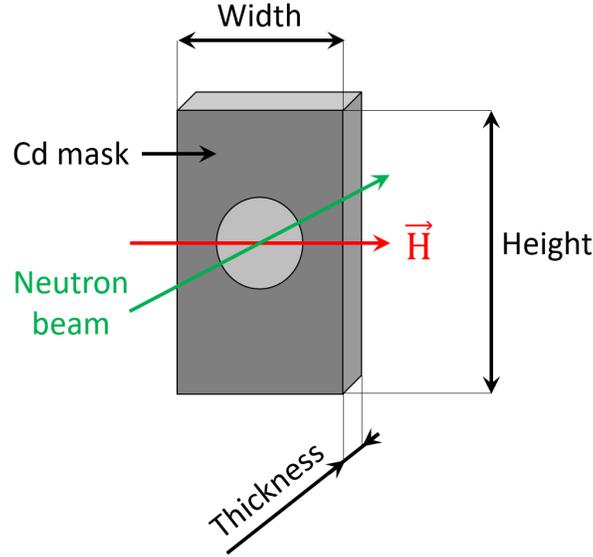}
\caption{\label{fig:sample_geometry}Experimental sample geometry used during the SANS experiment. The horizontal magnetic field $\vec{H}$ was applied transverse to the largest sample dimension.}
\end{center}
\end{figure}

For practical reasons, the magnetic field could not be applied along the largest dimensions of the samples (see Fig. \ref{fig:sample_geometry}). In order to obtain scaling laws depending on the \textit{internal} magnetic field $\mu_{\rm 0}H_{\rm int}$ experienced by the samples, we have calculated the demagnetizing field using magnetization data presented in Sec. \ref{sec:magnetic} (for which demagnetizing field was negligible) and the demagnetization factor for a very flat ellipsoid given by Osborn\cite{Osborn1945} (Eq. 2.24). The latter approximation is justified in view of the effective sample dimensions (Tab. \ref{tab:sample_thick_SANS}).

\begin{table}[!ht]
\caption{\label{tab:sample_thick_SANS}Dimensions of the a-FeMn samples used in the SANS study.}
%\begin{ruledtabular}
\begin{center}
\begin{tabular}{cccc}
\hline
$x$ & Thickness (mm) & Width (mm) & Height (mm) \\ 
0.22 & 3.0 & 9.1 & 30.9 \\
0.235 & 2.1 & 8.0 & 36.6 \\
0.247 & 2.3 & 8.9 & 38.0 \\
0.3 & 1.6 & 8.6 & 37.7 \\
0.32 & 1.5 & 9.0 & 21.6 \\
0.35 & 1.9 & 9.2 & 33.8 \\
0.41 & 4.9 & 8.7 & 33.0 \\
\hline
\end{tabular}
\end{center}
%\end{ruledtabular}
\end{table} 

\subsection{Data correction}

In order to get scattering cross sections in absolute units, we follow the usual data reduction procedure. First, the contribution from the environment and direct beam are removed using

\begin{equation}\label{eq:isub}
	I^{\rm sub}(Q) = \frac{I(Q) / t(Q) - \frac{t(Q)}{t_{\rm empty~cell}(0)} \cdot I_{\rm empty~cell}(Q) / t_{\rm empty~cell}(Q)}{\Omega(Q)} \quad ,
\end{equation}

where $I(Q)$, $t(Q)$ and $\Omega(Q)$ are the $Q$-dependent raw intensities, sample transmissions and solid angles subtended by the corresponding detector pixels\cite{Brulet2007}. In Eq. \ref{eq:isub}, the subscript "empty~cell" denotes a measurement performed using the same sample holder assembly as for the sample (including the Al foil).

This subtraction procedure is applied to the SANS from the samples and from a Ni single crystal. The latter is used to transform the observed intensities in absolute cross sections, according to

\begin{equation}\label{eq:ni_scaling}
	\sigma (Q) = \frac{I_{\rm a-FeMn}^{\rm sub}(Q) \cdot t_{\rm Ni}(0) \cdot d_{\rm Ni} \cdot e_{\rm Ni}}{\langle I_{\rm Ni}^{\rm sub}(Q) \rangle \cdot t_{\rm a-FeMn}(0) \cdot d_{\rm a-FeMn} \cdot e_{\rm a-FeMn}} \cdot \sigma_{\rm Ni}^{\rm inc} \quad ,
\end{equation}

where $t$, $d$ and $e$ denote transmission, atomic density (see Sec. \ref{sec:samples}) and sample thickness (see Tab. \ref{tab:sample_thick_SANS}), respectively, while $\sigma_{\rm Ni}^{\rm inc} = 5.2/4\pi$\,barn.sr$^{-1}$ is the incoherent scattering cross section of Ni ($\langle ... \rangle$ denotes the average over the detector surface, where incoherent scattering of Ni is expected to be flat).

\subsection{Effect of a cooling field on the SANS of the x = 0.22 sample}

As noted in seminal experimental\cite{Prejean1980} and theoretical\cite{Fert1980} studies of spin glasses, the anisotropy field maintaining the remanent magnetization in the direction of an initial applied field strongly depends of the elements composing the studied material. This comes from additional terms in the Ruderman-Kittel-Kasuya-Yosida (RKKY) interaction for atoms with large spin-orbit coupling (such as Au, Pt, \textit{etc.}), leading to Dzyaloshinskii-Moriya (DM) anisotropy of unidirectional type.  In SG and RSG, this DM anisotropy modifies the torque magnetization, the shape and position of the hysteresis cycle, and the effect of a cooling field.

When the DM term is small, an applied field causes the magnetization to rotate "rigidly" against the DM-induced domain anisotropy. On the other hand, when the DM term is strong, domain configurations may relax, yielding a distribution in the strength of the resulting frictional torque. In the former case, this leads to displaced narrow hysteresis loops under field cooling (FC) conditions while, in the latter case, one observe undisplaced broadened loops, independent of the cooling conditions.

Rotational magnetization measurements in a-FeMn ($x = 0.235$) by Goeckner \& Kouvel\cite{Goeckner1991} have revealed that this system is much less "rigid" than {\it e.g.}  Ni$_{1-x}$Mn$_{x}$, where the anisotropy field is much smaller\cite{Kouvel1987}. 

\begin{figure}[!ht]
\begin{center}
\includegraphics[width=\textwidth]{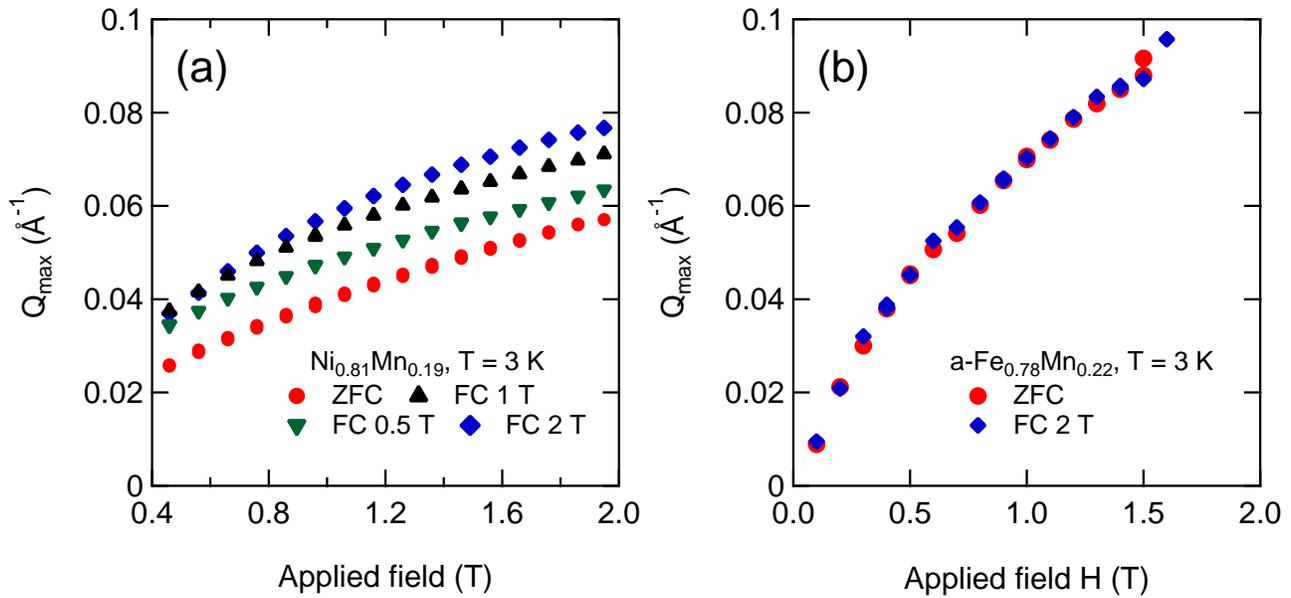}
\caption{\label{figS5}Field-dependence of the position $Q_{\rm max}$ of the maximum in transverse magnetic cross section $\sigma_{\rm T}(Q)$ for \textbf{(a)} Ni$_{0.81}$Mn$_{0.19}$ single crystal\cite{Mirebeau2018} and \textbf{(b)} a-Fe$_{0.78}$Mn$_{0.22}$.}
\end{center}
\end{figure}

To test the influence of the DM anisotropy on the vortex size, we have checked the effect of a cooling field on the scaling law of their average size {\it vs} field $Q_{\rm max} = f(\mu_{\rm 0}H_{\rm int})$ (see Eq. 3 of main text) in both systems\cite{Mirebeau2015}. The result is shown on Fig. \ref{figS5}, clearly demonstrating that $Q_{\rm max}$ is progressively shifted upwards by a cooling field in Ni$_{0.81}$Mn$_{0.19}$ while it remains unchanged in the case of a-Fe$_{0.78}$Mn$_{0.22}$. In other words, FC induces an extra magnetic field which reduces the vortex size in NiMn, whereas it has no effect in a-FeMn.  This shows that the DM anisotropy indeed plays a role on the vortex landscape, as it does on the magnetization, and underscores  the strong relation between the spin vortices and the underlying ferromagnetic vacuum.

%%%%%%%%%%%%%%%%%%%%%%%%%%%%%%
%%%%%%%%%%%%%%%%%%%%%%%%%%%%%%
%%%%%%%%%%%%%%%%%%%%%%%%%%%%%%
\clearpage
\newpage
\section{Monte Carlo simulations}
\label{sec:mcsims}
%%%%%%%%%%%%%%%%%%%%%%%%%%%%%%
%%%%%%%%%%%%%%%%%%%%%%%%%%%%%%
%%%%%%%%%%%%%%%%%%%%%%%%%%%%%%
\subsection{Model and parameters of the simulations}

Monte Carlo simulations presented in the main text have been performed using a standard local update algorithm on a square lattice containing $L \times L = 10^4$~spins, allowed to point in all directions of the 3d space ({\it i.e.} we work out a 2d Heisenberg model). "Impurity" spins are randomly spread over the matrix with the desired concentration $x$ to form a binary alloy with composition A$_{1-x}$B$_{x}$. These impurities emulate the Mn ions and are antiferromagnetically (AFM) coupled with other first neighbor impurity spins ($J = -1$), while all other couplings are ferromagnetic (FM, $J = +1$). Moments are taken to be equal for all spins, namely $M = 1$.

In order to obtain reasonable convergence times, we have used the "adaptative" algorithm proposed by Alzate-Cardona {\it et al.}\cite{Alzate-Cardona2019}. Trial moves of individual spin orientations are performed within a cone having an opening angle $\nu = 60^{\circ}$ with respect to the initial spin orientation. After each Monte Carlo step (MCS, corresponding to $10^4$ moves), $\nu$ is modified according to $\nu_{\rm new} \rightarrow \nu_{\rm old} \times f$ with 

\begin{equation}\label{eq:sigma_scalefactor}
	f = \frac{0.5}{1-R_{\rm old}} \quad ,
\end{equation}

where $R_{\rm old}$ is the acceptance rate observed during the previous MCS. This procedure allows keeping the average acceptance rate of the algorithm close to 50 \%, thereby leading to a relatively quick convergence. For each move, the classical energy

\begin{equation}\label{eq:classical_energy}
	\mathcal{H} = - \sum_{ij} J_{ij} \, \mathbf{S}_{i} \cdot \mathbf{S}_{j} - 0.672 \, H \sum_{i} S_{i}^{z}
\end{equation}

is calculated, where $\mathbf{S}_{i,j}$ are Heisenberg spins, $J_{ij}$ are random independent variables taking the value $\pm 1$ and the magnetic field $H$ is applied along the $z$ direction. The first sum in Eq. \ref{eq:classical_energy} runs over nearest-neighbor pairs. The factor 0.672 $(=\mu_{\rm B}/k_{\rm B})$ appearing in Eq. \ref{eq:classical_energy} allows getting energies in K for magnetic fields expressed in T. The acceptance of each trial move is tested against the Maxwell-Boltzmann statistics $\exp\left(-\Delta E / T\right)$, where $\Delta E$ is the energy difference between the "old" and "new" configurations). For each studied composition, we performed between 24 and 40 realizations of the following sequence:

\begin{enumerate}
	\item Random generation of the spin matrix at $T/J = 2$ and $H = 0$ with a chosen $c_{\rm imp}$,
	\item Slow cooling down to $T/J = 0.01$ at $H = 0$ (500 MCS per step),
	\item Field sweep in the $H/J = 0 - 1$ range, in small steps of $\Delta H/J = 0.01$ (500 MCS per step).
\end{enumerate}

%%%%%%%%%%%%%%%%%%%%%%%%%%%%%%
%%%%%%%%%%%%%%%%%%%%%%%%%%%%%%
%%%%%%%%%%%%%%%%%%%%%%%%%%%%%%
\subsection{Data analysis \& x-dependences}

In order to compare the result of our MC simulations with that of the SANS experiment, we compute the square of the Fourier transform of the spin matrices which is formally equivalent to the scattering cross section in the absence of correlation between defects (see Eq. 2 of main text). As in the experimental case, one can separate correlation functions for the transverse (T) and longitudinal (L) magnetization. Illustrative examples are shown in Fig. 4 of main text. Note that the explored momentum range is bounded downwards by the size of the spin maps ($100 \times 100$) and upwards by the nearest neighbor (NN) distance, but remains much larger than the one covered by SANS (see below).

\begin{figure}[!ht]
\begin{center}
\includegraphics[width=\textwidth]{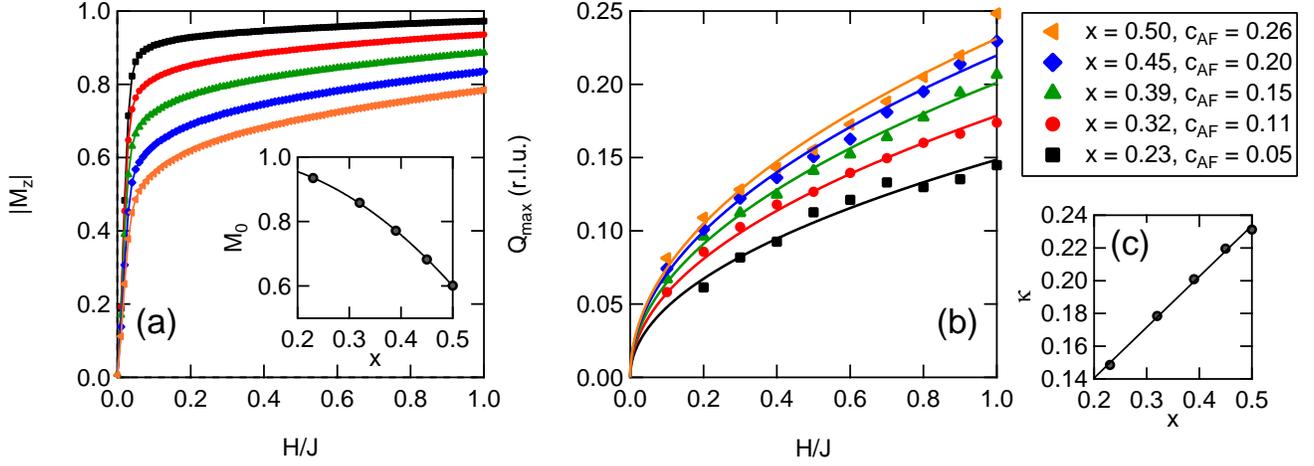}
\caption{\label{figS6}\textbf{(a)} $x$-dependence of the magnetization curves derived from the MC simulations. Inset shows the $x$-dependence of the spontaneous magnetization $M_{\rm 0}$. \textbf{(b)} Field-dependence of the position $Q_{\rm max}$ of the maximum in $\left|F_{\rm T}(Q)\right|^{2}$ for each studied compositions. Solid lines are results of a global fit of Eq. 3 of main text to the data. \textbf{(c)} $x$-dependence of the scaling parameter $\kappa$, extracted from a fit of Eq. 3 of main text to the data of panel \textbf{(b)}.}
\end{center}
\end{figure}

In addition to the $x = 0.23$ ($c_{\rm AFM} \approx 0.05$)-case discussed in the main text, we have also explored different concentrations in order to check the applicability of our primitive simulations when $x$ changes within the weakly frustrated side of the phase diagram ({\it i.e.}, for $\approx 0.05 \lesssim c_{\rm AFM} \lesssim 0.26$). The obtained magnetization curves are plotted in Fig. \ref{figS6}a, showing the same behavior as a function of increasing frustration as the experimental ones (Fig. 1b of main text and Fig. \ref{figS4}a of this supplement). We have also determined the scaling laws of $Q_{\rm max}$ as a function of the applied magnetic field (Fig. \ref{figS6}b). A global fit of Eq. 3 of main text to the data yields an exponent $\gamma = 0.49(1)$ that is slightly different from the experimental value ($\approx 0.39$). Of course, our simulations are only taking NN interactions on a square lattice into account. It is not surprising that the scaling laws are renormalized by the effect of an increased number of first neighbors (see Sec. \ref{sec:samples}) and longer-ranged interactions in the real amorphous metallic samples. However, the simulated scaling laws are very similar to the experimental ones in the sense that the scaling paramter $\kappa$, monitoring the "stiffness" of these curves, increases linearly with $x$ in agreement with the experiment (Fig. \ref{figS6}c). Taking these results together, we find that such a simplified model already captures many experimental features. Moreover, the MC simulations show that the scaling laws are indeed verified up to large concentrations of AF interactions, and to high values of the $H/J$ ratio. 

This simple model could of course be extended to account \textit{e.g.} for the behavior at the RSG-SG threshold, taking more realistic values for the moments, exchange constants and atomic connectivities.

\begin{figure}[!ht]
\begin{center}
\includegraphics[width=\textwidth]{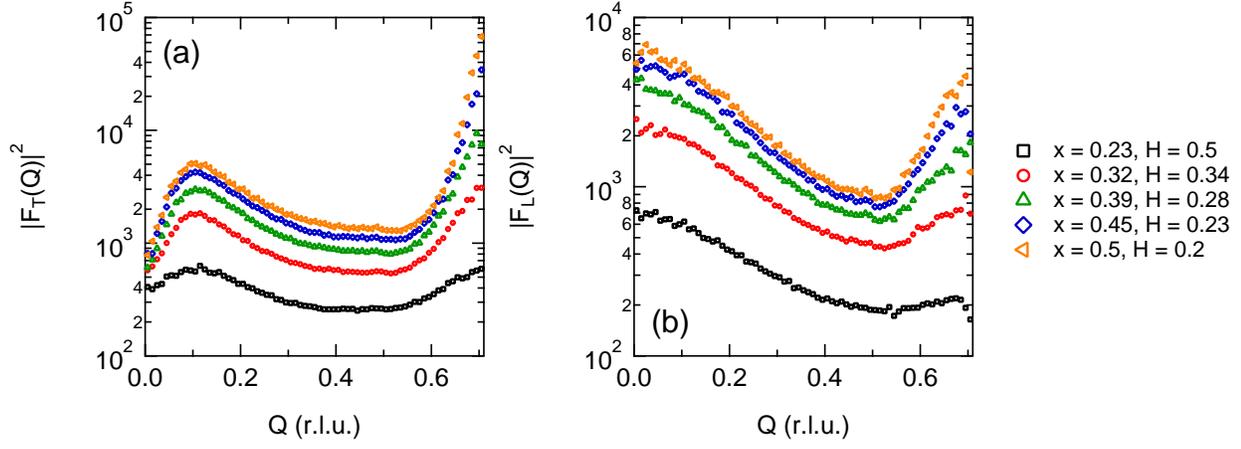}
\caption{\label{figS7}Squared Fourier transforms of the transverse \textbf{(a)} and longitudinal \textbf{(b)} spin correlations for different $x$ and magnetic field values, chosen to yield a constant $Q_{\rm max}$ (see text).}
\end{center}
\end{figure}

\newpage
As discussed above, the MC simulations allow extending the explored $Q$-range to values $Q \approx 1$, sensitive to the smallest interatomic distance. This is an interesting asset, since one can expect  in this Q-range (which naturally evades the SANS window), a growing AFM contribution to the scattering pattern as $x$ increases. As shown in Fig. \ref{figS7}, where we have selected field values such that $Q_{\rm max}$ stays constant for different $x$, this is indeed the case.

\begin{thebibliography}{42}
\expandafter\ifx\csname natexlab\endcsname\relax\def\natexlab#1{#1}\fi
\expandafter\ifx\csname bibnamefont\endcsname\relax
  \def\bibnamefont#1{#1}\fi
\expandafter\ifx\csname bibfnamefont\endcsname\relax
  \def\bibfnamefont#1{#1}\fi
\expandafter\ifx\csname citenamefont\endcsname\relax
  \def\citenamefont#1{#1}\fi
\expandafter\ifx\csname url\endcsname\relax
  \def\url#1{\texttt{#1}}\fi
\expandafter\ifx\csname urlprefix\endcsname\relax\def\urlprefix{URL }\fi
\providecommand{\bibinfo}[2]{#2}
\providecommand{\eprint}[2][]{\url{#2}}

\bibitem[{\citenamefont{Gabay and Toulouse}(1981)}]{Gabay1981}
\bibinfo{author}{\bibfnamefont{M.}~\bibnamefont{Gabay}} \bibnamefont{and}
  \bibinfo{author}{\bibfnamefont{G.}~\bibnamefont{Toulouse}},
  \bibinfo{journal}{Phys. Rev. Lett.} \textbf{\bibinfo{volume}{47}},
  \bibinfo{pages}{201} (\bibinfo{year}{1981}).

\bibitem[{\citenamefont{Imry and Ma}(1975)}]{Imry1975}
\bibinfo{author}{\bibfnamefont{Y.}~\bibnamefont{Imry}} \bibnamefont{and}
  \bibinfo{author}{\bibfnamefont{S.-k.} \bibnamefont{Ma}},
  \bibinfo{journal}{Phys. Rev. Lett.} \textbf{\bibinfo{volume}{35}},
  \bibinfo{pages}{1399} (\bibinfo{year}{1975}).

\bibitem[{\citenamefont{Aeppli et~al.}(1983)\citenamefont{Aeppli, Shapiro,
  Birgeneau, and Chen}}]{Aeppli1983}
\bibinfo{author}{\bibfnamefont{G.}~\bibnamefont{Aeppli}},
  \bibinfo{author}{\bibfnamefont{S.~M.} \bibnamefont{Shapiro}},
  \bibinfo{author}{\bibfnamefont{R.~J.} \bibnamefont{Birgeneau}},
  \bibnamefont{and} \bibinfo{author}{\bibfnamefont{H.~S.} \bibnamefont{Chen}},
  \bibinfo{journal}{Phys. Rev. B} \textbf{\bibinfo{volume}{28}},
  \bibinfo{pages}{5160} (\bibinfo{year}{1983}).

\bibitem[{\citenamefont{Niidera and Matsubara}(2007)}]{Niidera2007}
\bibinfo{author}{\bibfnamefont{S.}~\bibnamefont{Niidera}} \bibnamefont{and}
  \bibinfo{author}{\bibfnamefont{F.}~\bibnamefont{Matsubara}},
  \bibinfo{journal}{Phys. Rev. B} \textbf{\bibinfo{volume}{75}},
  \bibinfo{pages}{144413} (\bibinfo{year}{2007}).

\bibitem[{\citenamefont{Proctor et~al.}(2014)\citenamefont{Proctor, Garanin,
  and Chudnovsky}}]{Proctor2014}
\bibinfo{author}{\bibfnamefont{T.~C.} \bibnamefont{Proctor}},
  \bibinfo{author}{\bibfnamefont{D.~A.} \bibnamefont{Garanin}},
  \bibnamefont{and} \bibinfo{author}{\bibfnamefont{E.~M.}
  \bibnamefont{Chudnovsky}}, \bibinfo{journal}{Phys. Rev. Lett.}
  \textbf{\bibinfo{volume}{112}}, \bibinfo{pages}{097201}
  (\bibinfo{year}{2014}).

\bibitem[{\citenamefont{Chudnovsky and Garanin}(2018)}]{Chudnovsky2018}
\bibinfo{author}{\bibfnamefont{E.~M.} \bibnamefont{Chudnovsky}}
  \bibnamefont{and} \bibinfo{author}{\bibfnamefont{D.~A.}
  \bibnamefont{Garanin}}, \bibinfo{journal}{Phys. Rev. Lett.}
  \textbf{\bibinfo{volume}{121}}, \bibinfo{pages}{017201}
  (\bibinfo{year}{2018}).

\bibitem[{\citenamefont{Hennion et~al.}(1986)\citenamefont{Hennion, Mirebeau,
  Hennion, Lequien, and Hippert}}]{Hennion1986}
\bibinfo{author}{\bibfnamefont{M.}~\bibnamefont{Hennion}},
  \bibinfo{author}{\bibfnamefont{I.}~\bibnamefont{Mirebeau}},
  \bibinfo{author}{\bibfnamefont{B.}~\bibnamefont{Hennion}},
  \bibinfo{author}{\bibfnamefont{S.}~\bibnamefont{Lequien}}, \bibnamefont{and}
  \bibinfo{author}{\bibfnamefont{F.}~\bibnamefont{Hippert}},
  \bibinfo{journal}{EPL (Europhysics Letters)} \textbf{\bibinfo{volume}{2}},
  \bibinfo{pages}{393} (\bibinfo{year}{1986}).

\bibitem[{\citenamefont{B\"oni et~al.}(1986)\citenamefont{B\"oni, Shapiro, and
  Motoya}}]{Boeni1986}
\bibinfo{author}{\bibfnamefont{P.}~\bibnamefont{B\"oni}},
  \bibinfo{author}{\bibfnamefont{S.}~\bibnamefont{Shapiro}}, \bibnamefont{and}
  \bibinfo{author}{\bibfnamefont{K.}~\bibnamefont{Motoya}},
  \bibinfo{journal}{Solid State Communications} \textbf{\bibinfo{volume}{60}},
  \bibinfo{pages}{881 } (\bibinfo{year}{1986}).

\bibitem[{\citenamefont{Lequien et~al.}(1987)\citenamefont{Lequien, Mirebeau,
  Hennion, Hennion, Hippert, and Murani}}]{Lequien1987}
\bibinfo{author}{\bibfnamefont{S.}~\bibnamefont{Lequien}},
  \bibinfo{author}{\bibfnamefont{I.}~\bibnamefont{Mirebeau}},
  \bibinfo{author}{\bibfnamefont{M.}~\bibnamefont{Hennion}},
  \bibinfo{author}{\bibfnamefont{B.}~\bibnamefont{Hennion}},
  \bibinfo{author}{\bibfnamefont{F.}~\bibnamefont{Hippert}}, \bibnamefont{and}
  \bibinfo{author}{\bibfnamefont{A.~P.} \bibnamefont{Murani}},
  \bibinfo{journal}{Phys. Rev. B} \textbf{\bibinfo{volume}{35}},
  \bibinfo{pages}{7279} (\bibinfo{year}{1987}).

\bibitem[{\citenamefont{Hennion et~al.}(1988)\citenamefont{Hennion, Hennion,
  Mirebeau, Lequien, and Hippert}}]{Hennion1988}
\bibinfo{author}{\bibfnamefont{M.}~\bibnamefont{Hennion}},
  \bibinfo{author}{\bibfnamefont{B.}~\bibnamefont{Hennion}},
  \bibinfo{author}{\bibfnamefont{I.}~\bibnamefont{Mirebeau}},
  \bibinfo{author}{\bibfnamefont{S.}~\bibnamefont{Lequien}}, \bibnamefont{and}
  \bibinfo{author}{\bibfnamefont{F.}~\bibnamefont{Hippert}},
  \bibinfo{journal}{Journal of Applied Physics} \textbf{\bibinfo{volume}{63}},
  \bibinfo{pages}{4071} (\bibinfo{year}{1988}).

\bibitem[{\citenamefont{Kawamura and Tanemura}(1991)}]{Kawamura1991}
\bibinfo{author}{\bibfnamefont{H.}~\bibnamefont{Kawamura}} \bibnamefont{and}
  \bibinfo{author}{\bibfnamefont{M.}~\bibnamefont{Tanemura}},
  \bibinfo{journal}{Journal of the Physical Society of Japan}
  \textbf{\bibinfo{volume}{60}}, \bibinfo{pages}{1092} (\bibinfo{year}{1991}).

\bibitem[{\citenamefont{Yeshurun et~al.}(1981)\citenamefont{Yeshurun, Salamon,
  Rao, and Chen}}]{Yeshurun1981}
\bibinfo{author}{\bibfnamefont{Y.}~\bibnamefont{Yeshurun}},
  \bibinfo{author}{\bibfnamefont{M.~B.} \bibnamefont{Salamon}},
  \bibinfo{author}{\bibfnamefont{K.~V.} \bibnamefont{Rao}}, \bibnamefont{and}
  \bibinfo{author}{\bibfnamefont{H.~S.} \bibnamefont{Chen}},
  \bibinfo{journal}{Phys. Rev. B} \textbf{\bibinfo{volume}{24}},
  \bibinfo{pages}{1536} (\bibinfo{year}{1981}).

\bibitem[{\citenamefont{Mirebeau et~al.}(1997)\citenamefont{Mirebeau, Hennion,
  Gingras, Keren, Kojima, Larkin, Luke, Nachumi, Wu, Uemura
  et~al.}}]{Mirebeau1997}
\bibinfo{author}{\bibfnamefont{I.}~\bibnamefont{Mirebeau}},
  \bibinfo{author}{\bibfnamefont{M.}~\bibnamefont{Hennion}},
  \bibinfo{author}{\bibfnamefont{M.~J.~P.} \bibnamefont{Gingras}},
  \bibinfo{author}{\bibfnamefont{A.}~\bibnamefont{Keren}},
  \bibinfo{author}{\bibfnamefont{K.}~\bibnamefont{Kojima}},
  \bibinfo{author}{\bibfnamefont{M.}~\bibnamefont{Larkin}},
  \bibinfo{author}{\bibfnamefont{G.~M.} \bibnamefont{Luke}},
  \bibinfo{author}{\bibfnamefont{B.}~\bibnamefont{Nachumi}},
  \bibinfo{author}{\bibfnamefont{W.~D.} \bibnamefont{Wu}},
  \bibinfo{author}{\bibfnamefont{Y.~J.} \bibnamefont{Uemura}},
  \bibnamefont{et~al.}, \bibinfo{journal}{Hyperfine Interactions}
  \textbf{\bibinfo{volume}{104}}, \bibinfo{pages}{343} (\bibinfo{year}{1997}).

\bibitem[{\citenamefont{Salamon et~al.}(1980)\citenamefont{Salamon, Rao, and
  Chen}}]{Salamon1980}
\bibinfo{author}{\bibfnamefont{M.~B.} \bibnamefont{Salamon}},
  \bibinfo{author}{\bibfnamefont{K.~V.} \bibnamefont{Rao}}, \bibnamefont{and}
  \bibinfo{author}{\bibfnamefont{H.~S.} \bibnamefont{Chen}},
  \bibinfo{journal}{Phys. Rev. Lett.} \textbf{\bibinfo{volume}{44}},
  \bibinfo{pages}{596} (\bibinfo{year}{1980}).

\bibitem[{\citenamefont{Mirebeau et~al.}(1990)\citenamefont{Mirebeau, Itoh,
  Mitsuda, Watanabe, Endoh, Hennion, and Papoular}}]{Mirebeau1990}
\bibinfo{author}{\bibfnamefont{I.}~\bibnamefont{Mirebeau}},
  \bibinfo{author}{\bibfnamefont{S.}~\bibnamefont{Itoh}},
  \bibinfo{author}{\bibfnamefont{S.}~\bibnamefont{Mitsuda}},
  \bibinfo{author}{\bibfnamefont{T.}~\bibnamefont{Watanabe}},
  \bibinfo{author}{\bibfnamefont{Y.}~\bibnamefont{Endoh}},
  \bibinfo{author}{\bibfnamefont{M.}~\bibnamefont{Hennion}}, \bibnamefont{and}
  \bibinfo{author}{\bibfnamefont{R.}~\bibnamefont{Papoular}},
  \bibinfo{journal}{Phys. Rev. B} \textbf{\bibinfo{volume}{41}},
  \bibinfo{pages}{11405} (\bibinfo{year}{1990}).

\bibitem[{\citenamefont{Mirebeau}(1987)}]{Mirebeau1987}
\bibinfo{author}{\bibfnamefont{I.}~\bibnamefont{Mirebeau}}, Ph.D. thesis,
  \bibinfo{school}{Universit\'e de Paris-Sud} (\bibinfo{year}{1987}).

\bibitem[{\citenamefont{Mirebeau et~al.}(2018)\citenamefont{Mirebeau, Martin,
  Deutsch, Bannenberg, Pappas, Chaboussant, Cubitt, Decorse, and
  Leonov}}]{Mirebeau2018}
\bibinfo{author}{\bibfnamefont{I.}~\bibnamefont{Mirebeau}},
  \bibinfo{author}{\bibfnamefont{N.}~\bibnamefont{Martin}},
  \bibinfo{author}{\bibfnamefont{M.}~\bibnamefont{Deutsch}},
  \bibinfo{author}{\bibfnamefont{L.~J.} \bibnamefont{Bannenberg}},
  \bibinfo{author}{\bibfnamefont{C.}~\bibnamefont{Pappas}},
  \bibinfo{author}{\bibfnamefont{G.}~\bibnamefont{Chaboussant}},
  \bibinfo{author}{\bibfnamefont{R.}~\bibnamefont{Cubitt}},
  \bibinfo{author}{\bibfnamefont{C.}~\bibnamefont{Decorse}}, \bibnamefont{and}
  \bibinfo{author}{\bibfnamefont{A.~O.} \bibnamefont{Leonov}},
  \bibinfo{journal}{Phys. Rev. B} \textbf{\bibinfo{volume}{98}},
  \bibinfo{pages}{014420} (\bibinfo{year}{2018}).

\bibitem[{\citenamefont{Campbell and Gom{\`{e}}s}(1967)}]{Campbell1967}
\bibinfo{author}{\bibfnamefont{I.~A.} \bibnamefont{Campbell}} \bibnamefont{and}
  \bibinfo{author}{\bibfnamefont{A.~A.} \bibnamefont{Gom{\`{e}}s}},
  \bibinfo{journal}{Proceedings of the Physical Society}
  \textbf{\bibinfo{volume}{91}}, \bibinfo{pages}{319} (\bibinfo{year}{1967}).

\bibitem[{\citenamefont{Mirzoev et~al.}(2006)\citenamefont{Mirzoev, Yalalov,
  and Mirzaev}}]{Mirzoev2006}
\bibinfo{author}{\bibfnamefont{A.~A.} \bibnamefont{Mirzoev}},
  \bibinfo{author}{\bibfnamefont{M.~M.} \bibnamefont{Yalalov}},
  \bibnamefont{and} \bibinfo{author}{\bibfnamefont{D.~A.}
  \bibnamefont{Mirzaev}}, \bibinfo{journal}{The Physics of Metals and
  Metallography} \textbf{\bibinfo{volume}{101}}, \bibinfo{pages}{341}
  (\bibinfo{year}{2006}).

\bibitem[{\citenamefont{Schneider et~al.}(2018)\citenamefont{Schneider, Fu, and
  Barreteau}}]{Schneider2018}
\bibinfo{author}{\bibfnamefont{A.}~\bibnamefont{Schneider}},
  \bibinfo{author}{\bibfnamefont{C.-C.} \bibnamefont{Fu}}, \bibnamefont{and}
  \bibinfo{author}{\bibfnamefont{C.}~\bibnamefont{Barreteau}},
  \bibinfo{journal}{Phys. Rev. B} \textbf{\bibinfo{volume}{98}},
  \bibinfo{pages}{094426} (\bibinfo{year}{2018}).

\bibitem[{\citenamefont{Weissm\"uller et~al.}(1999)\citenamefont{Weissm\"uller,
  McMichael, Michels, and Shull}}]{Weissmuller1999}
\bibinfo{author}{\bibfnamefont{J.}~\bibnamefont{Weissm\"uller}},
  \bibinfo{author}{\bibfnamefont{R.}~\bibnamefont{McMichael}},
  \bibinfo{author}{\bibfnamefont{A.}~\bibnamefont{Michels}}, \bibnamefont{and}
  \bibinfo{author}{\bibfnamefont{R.}~\bibnamefont{Shull}}, \bibinfo{journal}{J.
  Res. Natl. Inst. Stand. Technol.} \textbf{\bibinfo{volume}{104}},
  \bibinfo{pages}{261} (\bibinfo{year}{1999}).

\bibitem[{\citenamefont{Mirebeau et~al.}(1986)\citenamefont{Mirebeau, Jehanno,
  Campbell, Hippert, Hennion, and Hennion}}]{Mirebeau1986}
\bibinfo{author}{\bibfnamefont{I.}~\bibnamefont{Mirebeau}},
  \bibinfo{author}{\bibfnamefont{G.}~\bibnamefont{Jehanno}},
  \bibinfo{author}{\bibfnamefont{I.}~\bibnamefont{Campbell}},
  \bibinfo{author}{\bibfnamefont{F.}~\bibnamefont{Hippert}},
  \bibinfo{author}{\bibfnamefont{B.}~\bibnamefont{Hennion}}, \bibnamefont{and}
  \bibinfo{author}{\bibfnamefont{M.}~\bibnamefont{Hennion}},
  \bibinfo{journal}{Journal of Magnetism and Magnetic Materials}
  \textbf{\bibinfo{volume}{54-57}}, \bibinfo{pages}{99} (\bibinfo{year}{1986}).

\bibitem[{\citenamefont{Senoussi et~al.}(1988)\citenamefont{Senoussi, Hadjoudj,
  and Fourmeaux}}]{Senoussi1988}
\bibinfo{author}{\bibfnamefont{S.}~\bibnamefont{Senoussi}},
  \bibinfo{author}{\bibfnamefont{S.}~\bibnamefont{Hadjoudj}}, \bibnamefont{and}
  \bibinfo{author}{\bibfnamefont{R.}~\bibnamefont{Fourmeaux}},
  \bibinfo{journal}{Phys. Rev. Lett.} \textbf{\bibinfo{volume}{61}},
  \bibinfo{pages}{1013} (\bibinfo{year}{1988}).

\bibitem[{\citenamefont{Kustov et~al.}(2017)\citenamefont{Kustov,
  Torrens-Serra, Salje, and Beshers}}]{Kustov2017}
\bibinfo{author}{\bibfnamefont{S.}~\bibnamefont{Kustov}},
  \bibinfo{author}{\bibfnamefont{J.}~\bibnamefont{Torrens-Serra}},
  \bibinfo{author}{\bibfnamefont{E.~K.~H.} \bibnamefont{Salje}},
  \bibnamefont{and} \bibinfo{author}{\bibfnamefont{D.~N.}
  \bibnamefont{Beshers}}, \bibinfo{journal}{Scientific Reports}
  \textbf{\bibinfo{volume}{7}}, \bibinfo{pages}{16846} (\bibinfo{year}{2017}).

\bibitem[{\citenamefont{{Toulouse, G.} and {Kl\'eman,
  M.}}(1976)}]{Toulouse1976}
\bibinfo{author}{\bibnamefont{{Toulouse, G.}}} \bibnamefont{and}
  \bibinfo{author}{\bibnamefont{{Kl\'eman, M.}}}, \bibinfo{journal}{J. Physique
  Lett.} \textbf{\bibinfo{volume}{37}}, \bibinfo{pages}{149}
  (\bibinfo{year}{1976}).

\bibitem[{\citenamefont{Kawamura}(2003)}]{Kawamura2003}
\bibinfo{author}{\bibfnamefont{H.}~\bibnamefont{Kawamura}},
  \bibinfo{journal}{Phys. Rev. Lett.} \textbf{\bibinfo{volume}{90}},
  \bibinfo{pages}{047202} (\bibinfo{year}{2003}).

\bibitem[{\citenamefont{Pureur et~al.}(2004)\citenamefont{Pureur, Fabris,
  Schaf, and Campbell}}]{Pureur2004}
\bibinfo{author}{\bibfnamefont{P.}~\bibnamefont{Pureur}},
  \bibinfo{author}{\bibfnamefont{F.~W.} \bibnamefont{Fabris}},
  \bibinfo{author}{\bibfnamefont{J.}~\bibnamefont{Schaf}}, \bibnamefont{and}
  \bibinfo{author}{\bibfnamefont{I.~A.} \bibnamefont{Campbell}},
  \bibinfo{journal}{EPL (Europhysics Letters)} \textbf{\bibinfo{volume}{67}},
  \bibinfo{pages}{123} (\bibinfo{year}{2004}).

\bibitem[{\citenamefont{Fabris et~al.}(2006)\citenamefont{Fabris, Pureur,
  Schaf, Vieira, and Campbell}}]{Fabris2006}
\bibinfo{author}{\bibfnamefont{F.~W.} \bibnamefont{Fabris}},
  \bibinfo{author}{\bibfnamefont{P.}~\bibnamefont{Pureur}},
  \bibinfo{author}{\bibfnamefont{J.}~\bibnamefont{Schaf}},
  \bibinfo{author}{\bibfnamefont{V.~N.} \bibnamefont{Vieira}},
  \bibnamefont{and} \bibinfo{author}{\bibfnamefont{I.~A.}
  \bibnamefont{Campbell}}, \bibinfo{journal}{Phys. Rev. B}
  \textbf{\bibinfo{volume}{74}}, \bibinfo{pages}{214201}
  (\bibinfo{year}{2006}).

\bibitem[{\citenamefont{Mirebeau et~al.}(2015)\citenamefont{Mirebeau,
  Bannenberg, Cubitt, Deutsch, Martin, and Pappas}}]{Mirebeau2015}
\bibinfo{author}{\bibfnamefont{I.}~\bibnamefont{Mirebeau}},
  \bibinfo{author}{\bibfnamefont{L.}~\bibnamefont{Bannenberg}},
  \bibinfo{author}{\bibfnamefont{R.}~\bibnamefont{Cubitt}},
  \bibinfo{author}{\bibfnamefont{M.}~\bibnamefont{Deutsch}},
  \bibinfo{author}{\bibfnamefont{N.}~\bibnamefont{Martin}}, \bibnamefont{and}
  \bibinfo{author}{\bibfnamefont{C.}~\bibnamefont{Pappas}}
  (\bibinfo{year}{2015}).

\bibitem[{\citenamefont{Alzate-Cardona
  et~al.}(2019)\citenamefont{Alzate-Cardona, Sabogal-Su{\'{a}}rez, Evans, and
  Restrepo-Parra}}]{Alzate-Cardona2019}
\bibinfo{author}{\bibfnamefont{J.~D.} \bibnamefont{Alzate-Cardona}},
  \bibinfo{author}{\bibfnamefont{D.}~\bibnamefont{Sabogal-Su{\'{a}}rez}},
  \bibinfo{author}{\bibfnamefont{R.~F.~L.} \bibnamefont{Evans}},
  \bibnamefont{and}
  \bibinfo{author}{\bibfnamefont{E.}~\bibnamefont{Restrepo-Parra}},
  \bibinfo{journal}{Journal of Physics: Condensed Matter}
  \textbf{\bibinfo{volume}{31}}, \bibinfo{pages}{095802}
  (\bibinfo{year}{2019}).

\bibitem[{\citenamefont{Bellissent et~al.}(1993)\citenamefont{Bellissent,
  Galli, Grinstaff, Migliardo, and Suslick}}]{Bellissent1993}
\bibinfo{author}{\bibfnamefont{R.}~\bibnamefont{Bellissent}},
  \bibinfo{author}{\bibfnamefont{G.}~\bibnamefont{Galli}},
  \bibinfo{author}{\bibfnamefont{M.~W.} \bibnamefont{Grinstaff}},
  \bibinfo{author}{\bibfnamefont{P.}~\bibnamefont{Migliardo}},
  \bibnamefont{and} \bibinfo{author}{\bibfnamefont{K.~S.}
  \bibnamefont{Suslick}}, \bibinfo{journal}{Phys. Rev. B}
  \textbf{\bibinfo{volume}{48}}, \bibinfo{pages}{15797} (\bibinfo{year}{1993}).

\bibitem[{\citenamefont{Waseda and Suzuki}(1970)}]{Waseda1970}
\bibinfo{author}{\bibfnamefont{Y.}~\bibnamefont{Waseda}} \bibnamefont{and}
  \bibinfo{author}{\bibfnamefont{K.}~\bibnamefont{Suzuki}},
  \bibinfo{journal}{physica status solidi (b)} \textbf{\bibinfo{volume}{39}},
  \bibinfo{pages}{669} (\bibinfo{year}{1970}).

\bibitem[{\citenamefont{Shull and Wilkinson}(1955)}]{Shull1955}
\bibinfo{author}{\bibfnamefont{C.~G.} \bibnamefont{Shull}} \bibnamefont{and}
  \bibinfo{author}{\bibfnamefont{M.~K.} \bibnamefont{Wilkinson}},
  \bibinfo{journal}{Phys. Rev.} \textbf{\bibinfo{volume}{97}},
  \bibinfo{pages}{304} (\bibinfo{year}{1955}).

\bibitem[{\citenamefont{Cowlam and Carr}(1985)}]{Cowlam1985}
\bibinfo{author}{\bibfnamefont{N.}~\bibnamefont{Cowlam}} \bibnamefont{and}
  \bibinfo{author}{\bibfnamefont{G.~E.} \bibnamefont{Carr}},
  \bibinfo{journal}{Journal of Physics F: Metal Physics}
  \textbf{\bibinfo{volume}{15}}, \bibinfo{pages}{1109} (\bibinfo{year}{1985}).

\bibitem[{\citenamefont{Grinstaff et~al.}(1993)\citenamefont{Grinstaff,
  Salamon, and Suslick}}]{Grinstaff1993}
\bibinfo{author}{\bibfnamefont{M.~W.} \bibnamefont{Grinstaff}},
  \bibinfo{author}{\bibfnamefont{M.~B.} \bibnamefont{Salamon}},
  \bibnamefont{and} \bibinfo{author}{\bibfnamefont{K.~S.}
  \bibnamefont{Suslick}}, \bibinfo{journal}{Phys. Rev. B}
  \textbf{\bibinfo{volume}{48}}, \bibinfo{pages}{269} (\bibinfo{year}{1993}).

\bibitem[{\citenamefont{Durand and Yung}(1977)}]{Durand1977}
\bibinfo{author}{\bibfnamefont{J.}~\bibnamefont{Durand}} \bibnamefont{and}
  \bibinfo{author}{\bibfnamefont{M.}~\bibnamefont{Yung}},
  \emph{\bibinfo{title}{Electronic and Magnetic Properties of Amorphous Fe-P-B
  Alloys}} (\bibinfo{publisher}{Springer US}, \bibinfo{address}{Boston, MA},
  \bibinfo{year}{1977}), pp. \bibinfo{pages}{275--288}.

\bibitem[{\citenamefont{Osborn}(1945)}]{Osborn1945}
\bibinfo{author}{\bibfnamefont{J.~A.} \bibnamefont{Osborn}},
  \bibinfo{journal}{Phys. Rev.} \textbf{\bibinfo{volume}{67}},
  \bibinfo{pages}{351} (\bibinfo{year}{1945}).

\bibitem[{\citenamefont{Br{\^{u}}let et~al.}(2007)\citenamefont{Br{\^{u}}let,
  Lairez, Lapp, and Cotton}}]{Brulet2007}
\bibinfo{author}{\bibfnamefont{A.}~\bibnamefont{Br{\^{u}}let}},
  \bibinfo{author}{\bibfnamefont{D.}~\bibnamefont{Lairez}},
  \bibinfo{author}{\bibfnamefont{A.}~\bibnamefont{Lapp}}, \bibnamefont{and}
  \bibinfo{author}{\bibfnamefont{J.-P.} \bibnamefont{Cotton}},
  \bibinfo{journal}{Journal of Applied Crystallography}
  \textbf{\bibinfo{volume}{40}}, \bibinfo{pages}{165} (\bibinfo{year}{2007}).

\bibitem[{\citenamefont{Pr\'ejean et~al.}(1980)\citenamefont{Pr\'ejean,
  Joliclerc, and Monod}}]{Prejean1980}
\bibinfo{author}{\bibfnamefont{J.}~\bibnamefont{Pr\'ejean}},
  \bibinfo{author}{\bibfnamefont{M.}~\bibnamefont{Joliclerc}},
  \bibnamefont{and} \bibinfo{author}{\bibfnamefont{P.}~\bibnamefont{Monod}},
  \bibinfo{journal}{J. Phys. France} \textbf{\bibinfo{volume}{41}},
  \bibinfo{pages}{427} (\bibinfo{year}{1980}).

\bibitem[{\citenamefont{Fert and Levy}(1980)}]{Fert1980}
\bibinfo{author}{\bibfnamefont{A.}~\bibnamefont{Fert}} \bibnamefont{and}
  \bibinfo{author}{\bibfnamefont{P.~M.} \bibnamefont{Levy}},
  \bibinfo{journal}{Phys. Rev. Lett.} \textbf{\bibinfo{volume}{44}},
  \bibinfo{pages}{1538} (\bibinfo{year}{1980}).

\bibitem[{\citenamefont{Goeckner and Kouvel}(1991)}]{Goeckner1991}
\bibinfo{author}{\bibfnamefont{H.}~\bibnamefont{Goeckner}} \bibnamefont{and}
  \bibinfo{author}{\bibfnamefont{J.}~\bibnamefont{Kouvel}},
  \bibinfo{journal}{Journal of Applied Physics} \textbf{\bibinfo{volume}{70}},
  \bibinfo{pages}{6089} (\bibinfo{year}{1991}).

\bibitem[{\citenamefont{Kouvel et~al.}(1987)\citenamefont{Kouvel, Abdul-Razzaq,
  and Ziq}}]{Kouvel1987}
\bibinfo{author}{\bibfnamefont{J.~S.} \bibnamefont{Kouvel}},
  \bibinfo{author}{\bibfnamefont{W.}~\bibnamefont{Abdul-Razzaq}},
  \bibnamefont{and} \bibinfo{author}{\bibfnamefont{K.}~\bibnamefont{Ziq}},
  \bibinfo{journal}{Phys. Rev. B} \textbf{\bibinfo{volume}{35}},
  \bibinfo{pages}{1768} (\bibinfo{year}{1987}).

\end{thebibliography}
\end{document}